\documentclass[prx,twocolumn,amsmath,amssymb,superscriptaddress,floatfix,nofootinbib,aps]{revtex4-2}
\usepackage{amsmath}
\usepackage{amsfonts}
\usepackage{graphicx}
\usepackage{dcolumn}
\usepackage{bm}
\usepackage{amsthm}
\usepackage[cmyk]{xcolor}
\usepackage[colorlinks,bookmarks=false,citecolor=magenta,linkcolor=magenta,urlcolor=magenta]{hyperref}
\usepackage{braket}
\usepackage{hyperref}
\usepackage{orcidlink}

\begin{document}
\preprint{APS/123-QED}


\title{Distinct many-body scars and emergent quantum phases driven by competing interactions in two-species Rydberg arrays}

\author{Lei-Yi-Nan Liu \orcidlink{0009-0005-3072-3650}}
\affiliation{
 School of Physics, Beihang University, Beijing 100191, China
}

\author{Shun-Yao Yu}
\affiliation{
 School of Physics, Beihang University, Beijing 100191, China
}

\author{Shi-Rong Peng}
\affiliation{
 School of Physics, Beihang University, Beijing 100191, China
}

\author{Jie Sheng}
\affiliation{
 School of Physics, Beihang University, Beijing 100191, China
}

\author{Su Yi}
\affiliation{CAS Key Laboratory of Theoretical Physics, Institute of 
Theoretical Physics, Chinese Academy of Sciences, Beijing 100190, China}
\affiliation{Peng Huanwu Collaborative Center for Research and Education, Beihang University, Beijing 100191, China}

\author{Peng Xu}
\email{etherxp@wipm.ac.cn}
\affiliation{State Key Laboratory of Magnetic Resonance and Atomic and Molecular Physics, Wuhan Institute of Physics and Mathematics, Innovation Academy for Precision Measurement Science and Technology, Chinese Academy of Sciences, Wuhan 430071, China}

\author{Shou-Shu Gong}
\email{shoushu.gong@gbu.edu.cn}
\affiliation{School of Physical Sciences, Great Bay University, Dongguan 523000, China, and 
Great Bay Institute for Advanced Study, Dongguan 523000, China
}

\author{Tao Shi}
\email{tshi@itp.ac.cn}
\affiliation{CAS Key Laboratory of Theoretical Physics, Institute of 
Theoretical Physics, Chinese Academy of Sciences, Beijing 100190, China}

\author{Jian Cui \orcidlink{0000-0001-6643-7625}}
\email{jiancui@buaa.edu.cn}
\affiliation{
 School of Physics, Beihang University, Beijing 100191, China
}

\date{\today}

\begin{abstract}
Rydberg atom arrays composed of multiple atomic species stand as a highly promising platform 
for quantum computation. However, the underlying physics of these systems as quantum many-body 
systems remains poorly understood, owing to the intricate competition between attractive and 
repulsive interactions, phenomena that entirely defy the Rydberg blockade mechanism. 
We systematically calculate the ground-state phase diagram of alternating two-species atom arrays 
and their quench dynamics. Our findings reveal several novel quantum states absent in 
traditional cold-atom platforms, such as the period-4 product state $\ket{1100...}$, the period-6 product state $\ket{111000...}$, and an order-disorder mixed phase. 
In the quench dynamics, we confirm $\mathbb{Z}_2$ ordered state qualify as novel 
quantum many-body scars. Based on our perturbation analysis, the underlying physics ought to be 
described by a series of Cooper pair states spanning the entire energy spectrum, rather than 
the PXP low-energy effective model. A detailed analysis is also provided regarding 
the experimental preparation of those product states. 
Numerical evidence demonstrates that the proposed scheme exhibits robustness against 
typical experimental imperfections, thereby confirming its experimental feasibility. 
Moreover, the ground-state problem of the two-species array naturally maps to more general combinatorial optimization problems, extending the class of optimization tasks accessible to programmable neutral-atom quantum processors. Our work paves a new way for quantum simulation of novel quantum many-body states, which emerge from the interplay between competing interactions among different atom species and quantum fluctuations.

\end{abstract}


\maketitle

\section{Introduction}
Neutral atom arrays have emerged as a promising platform 
offering unprecedented control and tunability over quantum 
interactions. 
Based on the long-range dipole-dipole interactions between 
atoms in Rydberg states \cite{Saffman200902} and current optical 
tweezers techniques, the neutral atom arrays are ideal 
candidates for exploring exotic quantum phenomena 
\cite{Browaeys2020,Lukin2019,Browaeys2018,Jaewook2024,Lukin2023}, 
creating high-fidelity quantum entangling gates \cite{Lukin202310,Saffman201609,Peng202204} 
and performing large-scale quantum computing tasks \cite{Lukin202402,Gerhard202402}. 
To fully take the advantage of neutral atoms, it is necessary to 
study the physical properties of neutral atom array. 
One pivotal property of Rydberg atoms is the 
so-called Rydberg blockade mechanism which is the direct result 
of strong interactions between neighbouring Rydberg atoms. 
The blockade mechanism preventing nearby atoms to be simultaneously 
excited to the Rydberg state leads to various interesting phases 
and unique dynamical properties in the Rydberg atom arrays.

There has been extensive studies on the phase diagram of Rydberg atom arrays. 
In one dimension, Rydberg atom arrays undergoes quantum phase transitions 
as the external parameters are adjusted, revealing distinct phases 
including a disordered phase, different translational symmetry breaking ordered phases 
as well as Luttinger liquid phase usually referred to as floating phase, 
all of which have been observed in experiments \cite{Lukin2016,Lukin201711,Jin2024}. 
Rydberg atoms can also be arranged in arbitrary geometries in two or three spatial dimensions. 
Together with the strong interaction between Rydberg atoms, it is possible to realize 
a wide range of interacting spin models with high-fidelity manipulation 
and may lead to various exotic quantum phenomena including spin liquids 
or other topologically ordered states and emergent gauge fields~\cite{Lukin202107,Sachdev202003,Sachdev2021,Zoller202006,Browaeys2019}. 

Besides static states, Rydberg atoms, as a leading platform for quantum simulation, 
have also been used to study the dynamical properties of 
quantum systems including Kibble-Zurek mechanism \cite{Lukin2019}, 
time crystal \cite{YouLi202407,LaRocca202001,BaoSenShi2024}, 
many-body localization \cite{Markus201510} and quantum many-body scars \cite{Lukin201711,Lukin2021}. 
The Rydberg atom array shows non-trivial oscillations of 
domain wall density after a sudden quench first observed in experiment \cite{Lukin201711} which 
seems to violate the eigenstates thermalization hypothesis (ETH). 
It was later understood that this phenomenon can be explained by the quantum many-body scars 
embedded within the eigenstates of the low-energy effective Hamiltonian of the Rydberg atom array, 
commonly known as the PXP model. Although the PXP model is intrinsically chaotic, 
it hosts a set of weakly entangled eigenstates with approximately equal energy spacing. 
These atypical eigenstates, named quantum many-body scars, enable the system to exhibit 
non-ergodic dynamics with certain initial states, despite the underlying chaotic nature of the model.
~\cite{Papic201810,Papic202106}. 
In the context of the PXP model, many methods have been developed to 
study the non-thermalizing behavior of the Rydberg atom array such as analyzing the 
quench dynamics in a parameterized phase space, using forward scattering approximation to 
construct the Krylov subspace and seeking the quasi-symmetry hidden in the system \cite{Lukin201901,Papic201807,Pratik2022}.

The intriguing range of phenomena observed in single-species Rydberg 
atom arrays is mostly based on the Rydberg blockade mechanism. 
However, the extent to which the potential of these arrays can be further 
harnessed to reveal new physical phenomenon beyond the blockade mechanism remains an open question. 
Therefore, in this paper we introduce a new degree of freedom 
to the Rydberg atom array, the competing interactions arising from different atom species. 
The two-species Rydberg atom array has already been realized experimentally and many applications of the multi-species Rydberg atom array has been proposed 
\cite{dual-species-array1,
dual-species-array4,dual-species-array2,dual-species-array-app,dual-species-array3,dual-species-array-app2}. 
The various interactions between different species of 
atoms lead to a richer phenomena in both phase diagram and quench dynamics. 
Unlike single-species Rydberg atom array where two Rydberg atom always repel each other, 
the two-species or multi-species Rydberg atom array can exhibit both attractive and 
repulsive interaction at the same time. These competing interactions together with 
external laser field may induce new phases that cannot be observed in single-species 
atom array. Here we first report novel ground-state configurations in this regard. 
In addition, the complex interactions in two-species atom array can also 
induce non-trivial dynamical phenomenon where we found the non-thermalizing behavior 
of the Loschmidt echo which can be explained by the quantum many-body scar (QMBS) states 
in the system spectrum similar to the case in the ``PXP'' model,
yet these scar states can be constructed using a set of basis 
composed of quasi-particle states derived in the mean-field framework. 
Unlike the PXP model, the scar states here span the whole energy spectrum 
of the original Hamiltonian.
The numerical results also indicate this system exhibits dynamical quantum 
phase transitions (DQPT), and we argue that there is a deep connection between 
DQPT and QMBS here, i.e., the DQPT is induced by the time evolution in the QMBS subspace.

To close the gap between theoretical study and experiment realization, 
we also consider the possible preparation of these newly found 
exotic ground states and their robustness against environment noise. 
The numerical simulations suggest we stand a good 
chance of observing these exotic ground states in a real experiment.
Furthermore, the ability to prepare and identify the ground states of the
two-species Rydberg atom array is also of direct relevance for quantum
computation and combinatorial optimization. It is well known that the ground
state problem of a single-species Rydberg atom array can be mapped to the
maximum independent set problem on a graph, which has already been
demonstrated experimentally \cite{Ebadi2022}. 
Compared to single-species arrays, which are restricted to purely repulsive constraints (or purely attractive) and therefore primarily encode MIS-type problems, the two-species setting introduces mixed attractive-repulsive couplings at the native Hamiltonian level. This significantly extends the class of
combinatorial optimization tasks that can be addressed using programmable
neutral-atom quantum processors, further highlighting the potential of
multi-species Rydberg arrays as a versatile platform for quantum computation.

\section{Model and Parameter Setup}
\begin{figure}[h]
    \includegraphics[scale=0.44]{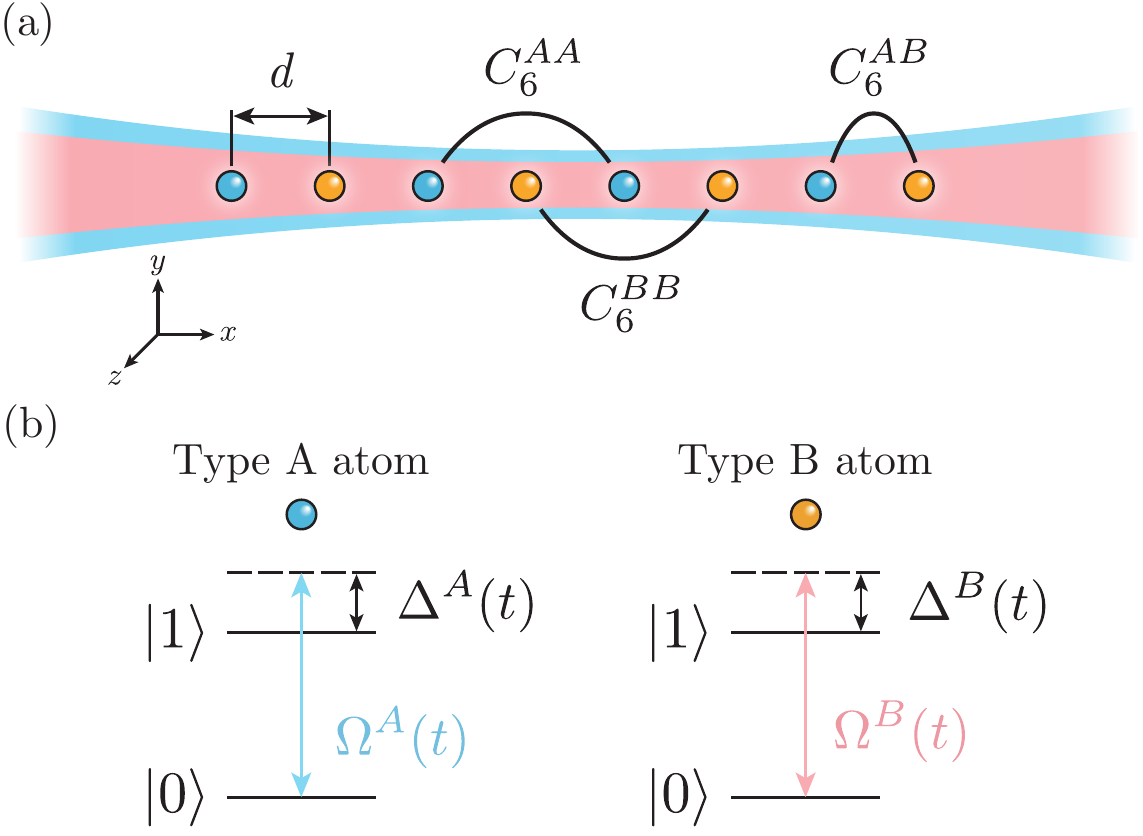}
    \caption{The two-species Rydberg atom array model. 
    (a) Two-species neutral atoms are alternatively arranged into one-dimensional 
    lattice with uniform spacing $d$ via optical tweezers where three kinds of interactions naturally exist. 
    (b) Illustration of energy levels, detunings and external fields of type A and B atoms.
    \label{model}
    }
\end{figure}

We study the two-species neutral atom arrays alternatively 
arranged into one-dimensional lattice with uniform spacing $d$ 
via optical tweezers. See Fig.~\ref{model}(a) where the atoms 
of type A(B) are sketched as blue(orange) circles. External laser 
fields uniformly shinning on those atoms can pump them between 
ground state labeled as $\ket{0}$ ($\ket{\circ}$) and certain 
Rydberg states labeled as $\ket{1}$ ($\ket{\bullet}$).
The system can be described by the following Hamiltonian
\begin{eqnarray}
    \hat{H}/\hbar &=& \hat{H}_0 + \hat{H}_1, \label{Hamiltonian} \\
    \hat{H}_0 &=& \sum_{s=A,B}\sum_{i\in s}\left(\frac{\Omega^s}{2}\hat{\sigma}_i^x - \Delta^s\hat{n}_i\right) \\
    \hat{H}_1 &=& \sum_{s=A,B}\sum_{s'=A,B}\sum_{i\in s,j\in s',i<j}\frac{C^{ss'}}{d^6}\frac{\hat{n}_i \hat{n}_j}{|i-j|^6} \label{H1}
\end{eqnarray}
where $\hat{\sigma}^x_i$ denotes the Pauli X operator and 
$\hat{n}_i = (\mathbb{I}+\hat{\sigma}^z_i)/2=\ket{1}\bra{1}$ 
is the excitation number operator. 
The parameter $s$ (also $s'$) stands for atom species. 
$\Omega^s$ as well as $\Delta^s$ depict the Rabi coupling and the frequency detuning of the external 
field acting on the type $s$ atom.
The term $\hat{H}_1$ accounts for the van der Waals interactions 
when both atoms are in excited Rydberg states 
where $C_6^{ss'}$ are determined by the species of 
the atoms as well as the selected Rydberg states. 
Namely, for the two-species atom arrays there exists 
three possible values $C_6^{AA}$, $C_6^{BB}$, $C_6^{AB}(=C_6^{BA})$ for the 
interactions, two possible transverse fields $\Omega^A(t)$, $\Omega^B(t)$ 
and two possible longitudinal fields $\Delta^A(t)$ and $\Delta^B(t)$ in general. 
For simplicity, in the this paper we assume that all atoms 
share equal coupling strength and equal detuning denoted as $\Omega$ and $\Delta$, respectively. 

\section{Ground state phase diagram}
The phase diagram of one-species Rydberg atom array has been throughly 
studied \cite{Lukin201711,Lukin2019,Ryd1-phase-diagram,floating-phase-numerical}. 
In the classical limit $\Omega = 0$, the ground states form 
a series of Rydberg crystal states and these transitional symmetry 
breaking states can be characterized by 
two coprime integer $p$ and $q$, where $q$ denotes 
the size of the unit cell and $p$ the number of 
Rydberg-state atoms in each unit cell. 
When quantum fluctuation is taken into 
consideration, i.e., $\Omega\neq 0$, the phase diagram 
still exhibits transitional symmetry breaking 
crystalline ground states denoted $\mathbb{Z}_2, \mathbb{Z}_3, \cdots, \mathbb{Z}_q$ 
with excitation density $1/2, 1/3, \cdots, 1/q$, 
and other area of the phase diagram is mainly dominated 
by the disordered phase \cite{floating-phase-numerical,Lukin201711}. 
Besides the disordered and crystalline states, the so-called 
floating phase lives between them which was predicted nummerically earlier and observed in 
experiment until recently \cite{floating-phase-numerical,Jin2024}.

The phase diagram is more complicated when we consider 
two-species atom arrays. 
In one-species atom array, the interaction 
decays rapidly ($\sim 1/r^6$) and the phase diagram 
can be understood using the 
blockade mechanism. However, in two-species atom array, 
the interaction certainly will not 
decay monotonically and in some cases the interactions 
strength between nearest neighbour (NN) $C_6^{AB}/d^6$ 
and next nearest neighbour (NNN) $C_6^{AA}/(2d)^6, C_6^{BB}/(2d)^6$
atoms can be very close thus leading to competitions of different configurations which 
may give birth to novel ground states that cannot be 
observed in one-species Rydberg atom arrays. 
Theoretically, 
there has been some investigations about the Ising model 
with nearest and next nearest neighbour interactions, i.e., the 
axial next-nearest neighbor Ising (ANNNI) model whose Hamiltonian reads
\begin{eqnarray}
    \hat{H}_{\mathrm{ANNNI}} &=& -J_1\sum_i\hat{\sigma}^z_i\hat{\sigma}^z_{i+1} - J_2\sum_i
    \hat{\sigma}^z_i\hat{\sigma}^z_{i+2} \nonumber\\
    & &- B\sum_i\hat{\sigma}^x_i, 
\end{eqnarray}
We further denote $h = B/J_1$ and the frustration parameter $\kappa = -J_2/J_1$ \cite{ANNNI-model2}. 
When $\kappa >1/2$ and $h = 0$, this model exhibits a so 
called anti-phase where the ground-state has a period of $4$ with two down-spins 
following two up-spins ($\ket{\uparrow\uparrow\downarrow\downarrow}$). If one turns on the 
transverse field ($h\neq0$), the floating phase can be found between the paramagnetic phase and 
the anti-phase \cite{ANNNI-model1,ANNNI-model2,ANNNI-model3}. 
However, the above theoretical predictions have never been found to exist in any materials 
or any quantum simulation platforms.

Due to the Ising-like interaction and the competition 
of NN and NNN interactions, we 
anticipate that the anti-phase can be found in the 
two-species Rydberg atom array, and indeed after extensive 
calculations of atoms' real energy potentials we found 
there are some $C_6$ coefficients that can induce 
the anti-phase ground state predicted by the ANNNI model,  
and moreover, we find several more versatile phases beyond the description of the ANNNI model.

\subsection{Classical limit}
\begin{table}[b]
    \caption{\label{state_configs}
    We list a few possible state configurations and their notations appeared in the main text and Appendix A for collation. 
    The configurations listed here can be regarded under periodic boundary condition or in a infinite system size and we 
    use $\ket{0}=\ket{\circ}$ and $\ket{1}=\ket{\bullet}$ to represent ground and excited state. 
    The first atom of each following configuration is always A atom just like in Eq.~\eqref{eq1100} and \eqref{eq111000}. } 
    \begin{ruledtabular}
    \begin{tabular}{ccccc}
    Notation & Configuration & $q$ & $p$ & $n$ \\
    \hline
    $\ket{\mathbb{D}}_A\ket{\mathbb{D}}_B$ & $|\bullet\bullet\bullet\bullet\bullet\bullet\bullet\bullet\bullet\bullet\bullet\bullet\rangle$ & 2 & 2 & 1 \\
    $\ket{\mathbb{Z}_2}_A\ket{\mathbb{Z}_2}_B$ & $|\bullet\bullet\circ\circ\bullet\bullet\circ\circ\bullet\bullet\circ\circ\rangle$ & 4 & 2 & 1/2 \\
    $\ket{\mathbb{Z}_3}_A\ket{\mathbb{Z}_3}_B$ & $|\bullet\bullet\circ\circ\circ\circ\bullet\bullet\circ\circ\circ\circ\rangle$ & 6 & 2 & 1/3 \\
    $\ket{\mathbb{Z}_4}_A\ket{\mathbb{Z}_4}_B$ & $|\bullet\bullet\circ\circ\circ\circ\circ\circ\bullet\bullet\circ\circ\circ\circ\circ\circ\rangle$ & 8 & 2 & 1/4 \\
    $\ket{\mathbb{Z}_2}_A\ket{\mathbb{D}}_B$ & $|\bullet\bullet\circ\bullet\bullet\bullet\circ\bullet\bullet\bullet\circ\bullet\rangle$ & 4 & 3 & 3/4\\
    $\ket{\mathbb{Z}_3}_A\ket{\mathbb{D}}_B$ & $|\bullet\bullet\circ\bullet\circ\bullet\bullet\bullet\circ\bullet\circ\bullet\rangle$ & 6 & 4 & 2/3 \\
    $\ket{\mathbb{D}}_A\ket{\mathbb{Z}_2}_B$ & $|\bullet\bullet\bullet\circ\bullet\bullet\bullet\circ\bullet\bullet\bullet\circ\rangle$ & 4 & 3 & 3/4\\
    $\ket{\mathbb{D}}_A\ket{\mathbb{Z}_3}_B$ & $|\bullet\bullet\bullet\circ\bullet\circ\bullet\bullet\bullet\circ\bullet\circ\rangle$ & 6 & 4 & 2/3 \\
    $\ket{\mathbb{Z}_3}_A\ket{\mathbb{Z}_3^2}_B$ & $|\bullet\bullet\circ\circ\circ\bullet\bullet\bullet\circ\circ\circ\bullet\rangle$ & 6 & 3 & 1/2 \\
    $\ket{\mathbb{Z}_3^2}_A\ket{\mathbb{Z}_3}_B$ & $|\bullet\bullet\bullet\circ\circ\circ\bullet\bullet\bullet\circ\circ\circ\rangle$ & 6 & 3 & 1/2 \\
    $\ket{\mathbb{Z}_3^2}_A\ket{\mathbb{D}}_B$ & $|\bullet\bullet\bullet\bullet\circ\bullet\bullet\bullet\bullet\bullet\circ\bullet\rangle$ & 6 & 5 & 5/6 \\
    $\ket{\mathbb{D}}_A\ket{\mathbb{Z}_3^2}_B$ & $|\bullet\bullet\bullet\bullet\bullet\circ\bullet\bullet\bullet\bullet\bullet\circ\rangle$ & 6 & 5 & 5/6 \\
    $\ket{\mathbb{Z}_3}_A\ket{\mathbb{Z}_2}_B$ & $|\bullet\bullet\circ\circ\circ\bullet\bullet\circ\circ\bullet\circ\circ\rangle$ & 12 & 5 & 5/12 \\
    $\ket{\mathbb{Z}_5^4}_A\ket{\mathbb{Z}_5^2}_B$ & 
    $|\bullet\bullet\bullet\circ\bullet\bullet\bullet\circ\circ\circ\rangle$ & 10 & 6 & 3/5
    \end{tabular}
    \end{ruledtabular}
\end{table}

\begin{figure}
    \centering
    \includegraphics[scale=0.39]{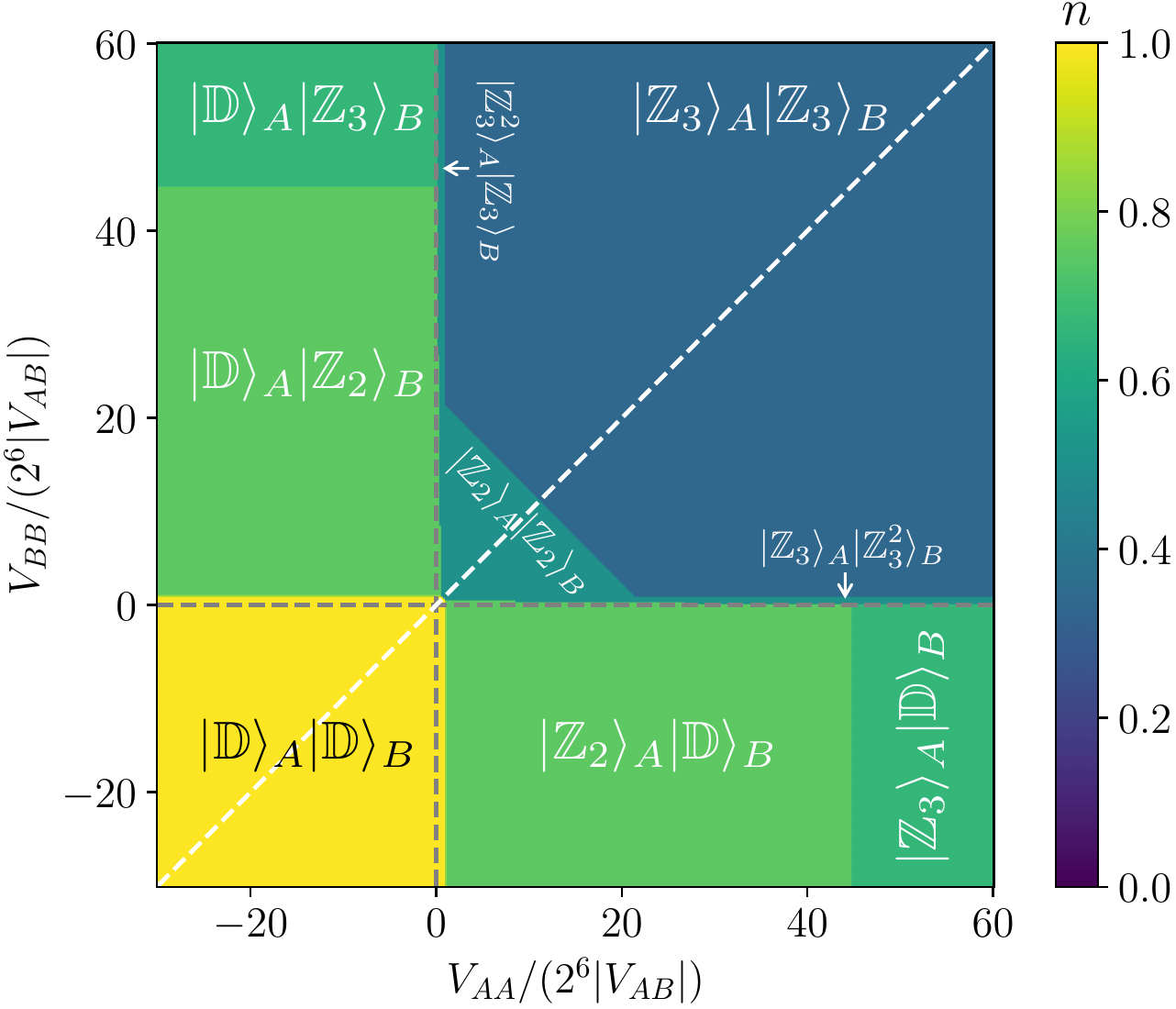}
    \caption{Classical phase diagram of Hamiltonian $\hat{H}_1$ in Eq.~\eqref{H1} 
    with $V_{AB}\equiv C_6^{AB}/d^6=-1$ the energy unit and 
    $V_{AA}\equiv C_6^{AA}/d^6, V_{BB} \equiv C_6^{BB}/d^6$ here. 
    The whole phase diagram is symmetric about the $V_{AA}=V_{BB}$ white dashed line. 
    The excitation density $n$ serves as the order parameter to distinguish different phases.  
    The grey dashed lines correspond to $V_{AA}=0$ and $V_{BB}=0$ respectively. 
    \label{classical_PD}}
\end{figure}

In the classical limit $\Omega = 0$, only $\Delta$ 
and the interaction strengths $C_6^{AA}$, $C_6^{BB}$, $C_6^{AB}$ 
dominate the Rydberg atom array. 
In the $\ket0, \ket1$ basis, all eigenstates of the system Hamiltonian
are product states so we use $\ket{\mathbb{Z}_{p}^{q}}$ ($\ket{\mathbb{D}}$)
to label different ordered (disordered) product states in sublattice of A and B atoms. 
For example, we use  $\ket{\mathbb{Z}_2}_A\ket{\mathbb{Z}_2}_B$ and 
$\ket{\mathbb{Z}_3^2}_A\ket{\mathbb{Z}_3}_B$ to denote $\ket{11001100\cdots}$ 
and $\ket{111000111000\cdots}$ respectively. 
Their interpretation is shown below 
\begin{eqnarray}
    \ket{\mathbb{Z}_2}_A\ket{\mathbb{Z}_2}_B &=& \ket{10}_A^{\otimes k}\ket{10}_B^{\otimes k}
    = \ket{1_A1_B0_A0_B}^{\otimes k} \label{eq1100}, \\
    \ket{\mathbb{Z}_3^2}_A\ket{\mathbb{Z}_3}_B 
    &=&\ket{110}_A^{\otimes k}\ket{100}_B^{\otimes k} = \ket{1_A1_B1_A0_B0_A0_B}^{\otimes k}\label{eq111000},\quad
\end{eqnarray}
where $\ket{\mathbb{Z}_q^p}$ stands for a state with unit cell length $q$ and excitation number $p$ and 
we use $\ket{\mathbb{Z}_q} = \ket{\mathbb{Z}_q^1}$ for simplicity.
To find proper Rydberg states that exactly produce 
$\ket{\mathbb{Z}_2}_A\ket{\mathbb{Z}_2}_B=\ket{11001100\cdots}$ and $\ket{\mathbb{Z}_3^2}_A\ket{\mathbb{Z}_3}_B=\ket{111000111000\cdots}$ states, we 
calculate the classical phase diagram when $\Delta = 0$ as shown in Fig.~\ref{classical_PD} 
where the meaning of all these phase labels can be found in Table~\ref{state_configs}. 
When the Hamiltonian parameters satisfy some particular conditions (see Fig.~\ref{classical_PD}), 
we then obatin ground states with period-4 pattern 
such as $\ket{\mathbb{Z}_2}_A\ket{\mathbb{Z}_2}_B$ which is the abovementioned anti-phase and  
period-6 pattern such as $\ket{\mathbb{Z}_3^2}_A\ket{\mathbb{Z}_3}_B$. 
We have found some interaction strengths ($C_6$ coefficients) of real atoms, 
for example, Rb and Cs or $^{85}$Rb and $^{87}$Rb 
with corresponding $\Delta$ revealing period-4 and period-6 ground states. 
These data are shown in Table \ref{interaction-data} which 
are selected from the interaction strength calculated from 
python package ARC \cite{ARC-python-package}. 
A more systematic analysis on 
the classical phase diagram are provided in Appendix A. 

\begin{table}[b]
    \caption{\label{interaction-data}
    Hamiltonian parameters revealing period-4 $\ket{1100\cdots}$ and 
    period-6 $\ket{111000\cdots}$ ground states. 
    $(n_1, n_2)$ is the principal quantum number of two atoms. 
    The unit of $C_6$ coefficients is $\mathrm{GHz}\cdot \mu \mathrm{m}^6$. 
    All the coefficents are calculated from the dipole-dipole 
    interactions between two atoms.
    The first(second) row gives rise to the $\ket{1100\cdots}$($\ket{111000\cdots}$) ground states 
    attached with the Rydberg state $\ket{88S_{1/2}}_{\mathrm{Rb}}\ket{86S_{1/2}}_{\mathrm{Cs}}$
    ($\ket{52S_{1/2}}_{\mathrm{^{85}Rb}}\ket{80S_{1/2}}_{\mathrm{^{87}Rb}}$). 
    All the interaction data are calculated via the python package ARC \cite{ARC-python-package}.} 
    \begin{ruledtabular}
    \begin{tabular}{ccccc}
    $(n_1, n_2)$ & $C^{AA}_6$ & $C^{BB}_6$ & $C^{AB}_6$ & $\Delta\cdot d^6 / \mathrm{GHz\cdot\mu m^6}$\\
    \hline
    $(88, 86)$ & 12696.34 & 7537.29 & $-131.93$ & $[-46.875,93.75]$ \\
    $(52, 80)$ & 24.83 & 4161.55 & $-2.16$ & $[-1.23, 0.04]$
    \end{tabular}
    \end{ruledtabular}
\end{table}

\subsection{Quantum Phase Diagram}

\begin{figure*}
    \centering
    \includegraphics[scale=0.595]{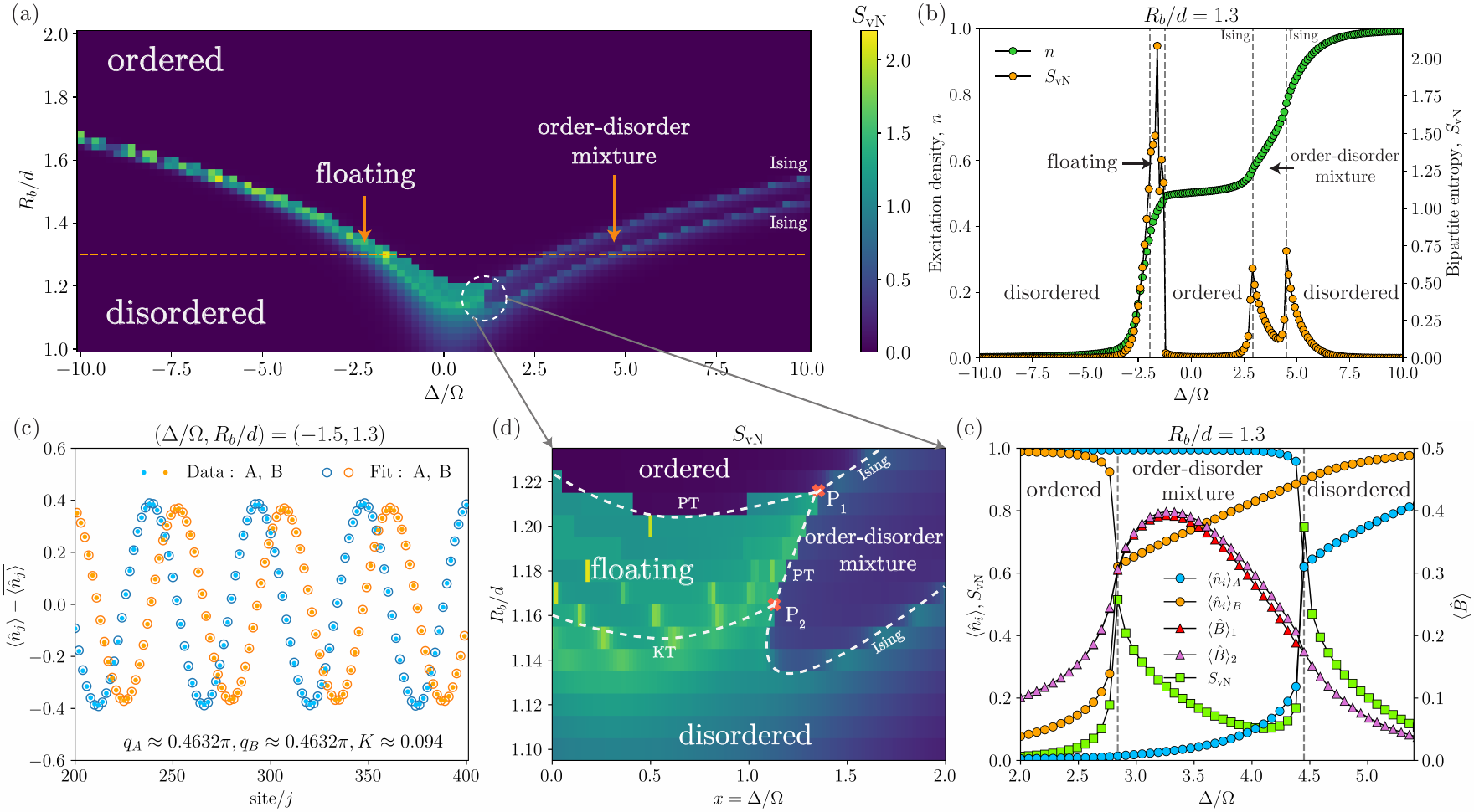}
    \caption{Phase diagram of Hamiltonian~\eqref{Hamiltonian} with $C_6$ coefficients from the first line of Table~\ref{interaction-data}, calculated using DMRG with system size $N = 282$.
    (a) The bipartite entanglement entropy $S_{\mathrm{vN}}$ reveals four distinct phases 
    in the $(\Delta/\Omega, R_b/d)$ plane: disordered, floating, 
    half-filling ordered ($\ket{1100\cdots}$), and an order-disorder mixture phase.
    (b) Excitation density $n$ and $S_{\mathrm{vN}}$ along the cut $R_b/d = 1.3$ (orange dashed line) 
    identify the order-disorder mixture phase in the range $2.8 \lesssim \Delta/\Omega \lesssim 4.5$. 
    (c) Friedel oscillations of the local excitation density in the floating phase, 
    induced by open boundary conditions, are fitted using Eq.~\eqref{friedel-fit} in a larger system (N=602). 
    Blue and orange dots represent A and B sublattice data; circles show the fits. 
    To minimize edge effects, fitting is restricted to $j \in [200, 400]$.
    %
    (d) Zoom-in of the phase diagram. Two multicritical points, labeled $P_1$ and $P_2$, are identified. Point $P_2$ is conventionally referred to as the Lifshitz point. The types of phase transitions are marked: PT for Pokrovsky–Talapov transition, KT/Ising for Kosterlitz–Thouless/Ising universality class.
    (e) Excitation densities on the two sublattices exhibit undergo transitions at different $\Delta/\Omega$ values, 
    indicating a novel order-disorder mixture phase in which translational symmetry breaks only on the A sublattice. 
    The bond energy $\langle \hat{B} \rangle$ between neighboring B atoms serves as an order parameter for this phase. 
    A small difference between $\langle \hat{B} \rangle_1$ and $\langle \hat{B} \rangle_2$ indicates 
    weak symmetry breaking in the B sublattice, induced by the strong one in the A sublattice.
    \label{phase-diagram}}
\end{figure*}

Now we mainly focus on the ground-state phase diagram of the system with the parameters shown in the first row of Table \ref{interaction-data}.
There are two free tuning parameters, which are chosen as $\Delta/\Omega$ and $R_b/d$ ($R_b \equiv \sqrt[6]{|C_{AB}|/\Omega}$) 
and $\Omega=1$ is fixed in the following numerical calculations. 
We perform density matrix renormalization group (DMRG)~\cite{DMRG1, ITensor, ITensor-r0.3} calculations on system size $N=282$ with open boundary conditions (OBC), where $N = 4n+2$ ($n \in \mathbb{N}^{+}$) can fit the particle distributions under OBC. 
In DMRG calculations, we consider the long-range interactions up to eight atom spacings. 
We keep the bond dimensions up to $512$, giving accurate results with the truncation error about $10^{-10}$. 

To map out the phase diagram, we first calculate the bipartite entanglement entropy $S_{\mathrm{vN}} = -\mathrm{Tr}\rho_{A}\log_2 \rho_{A}$, where $\rho_A$ is the reduced density matrix of the subsystem in the middle. 
The results are shown in Fig.~\ref{phase-diagram}(a).
Similar to the one-species system, the state in the small $R_b$ regime 
mainly depends on $\Delta/\Omega$ and we denote these states as disordered states in terms of no transitional symmetry breaking.
With growing interactions, the ordered $\ket{1100\cdots}$ state breaking translational symmetry dominates the phase diagram.
Besides, we identify a floating phase and an order-disorder mixture phase with divergent and enhanced entanglement entropy, respectively.

To explicitly demonstrate the quantum phases, we scan along the parameter line with fixed $R_b / d = 1.3$, as shown in Fig.~\ref{phase-diagram}(b), which characterizes the phases by excitation density $n$ and entropy.
We first discuss the floating phase, which can be well described by the Luttinger-liquid theory. The power-law decay of correlation functions in this phase is controlled by the Luttinger parameter 
$K$ \cite{Haldane1981,QuantumPhysicsin1D}.
Due to the gapless nature, the floating phase can hardly be identified in a small system, 
thus we carefully examine the DMRG results.
The open boundary in DMRG calculations acts as an impurity that can lead to the Friedel oscillation of excitation density. 
According to Boundary Conformal Field Theory (BCFT), the density profile generally can follow the behavior \cite{ANNNI-model2, Friedel-Fit1, Friedel-Fit2}
\begin{eqnarray}
    \langle\hat{n}_j\rangle &=& \overline{\langle \hat{n}_j\rangle}+(-1)^{\lfloor j/2\rfloor}\widetilde{\langle \hat{n}_j\rangle}, \\
    \widetilde{\langle \hat{n}_j\rangle}&\propto& \frac{\cos(qj+\alpha)}{[(N/\pi)\sin(\pi j/N)]^K}, \label{friedel-fit}
\end{eqnarray}
where $K$ is the Luttinger parameter, 
$q$ is the wave-vector that can be incommensurate, and $\alpha$ is the phase shift. 
However, the unit cell of our system is composed of two different atoms. 
Therefore, in principle we should regard this system as two majorana fermion fields (critical Ising chains) coupling with each other, which together behaves as a Dirac fermion field characterizing the low-energy bosonic excitations in the combined system. 
The low-energy effective bosonic field Hamiltonian is 
\begin{eqnarray}
    H = \displaystyle\frac{v}{2}\int\mathrm{d}x\left[
        K\Pi^2 + \frac{1}{K}\left(\partial_x\phi\right)^2
    \right],
\end{eqnarray}
where $\Pi$ and $\phi$ are canonical momentum and position,  
$v$ is the velocity of the corresponding gapless low-energy excitations \cite{Friedel-Fit2, future-work}.

The Luttinger parameters $K$ and the incommensurate wavevectors $q_A$ and $q_B$ 
are extracted at $(\Delta/\Omega, R_b/d) = (-1.5, 1.3)$ by independently fitting 
the Friedel oscillations on the two sublattices. According to bosonization theory, 
one expects $q_A = q_B$, and our results confirm this with $q_A \approx q_B \approx 0.4632\pi$. 
In the fitting process, we impose the constraint that the Luttinger parameter 
$K$ is the same on both sublattices and obatain $K\approx 0.094$. To verify this result, we also fit the 
correlation function defined in Eq.~\eqref{nn_correlation} \cite{QuantumPhysicsin1D,Karrasch2012} using data 
from variational uniform matrix product states (VUMPS) simulation \cite{VUMPS1}. 
\begin{equation}
    \langle \hat{n}_i \hat{n}_{i+x}\rangle-\langle \hat{n}_i\rangle\langle \hat{n}_{i+x}\rangle = \frac{K}{2\pi^2 x^2} + C\frac{\cos2qx}{x^{2K}}+\cdots
    \label{nn_correlation}
\end{equation}
The direct fitting of correlation functions yields $K\approx0.095$ and $q_A\approx q_B\approx 0.4615\pi$, which is 
consistent with the one obtained from DMRG simulations, within the fitting error.

We further characterize the gapless nature of the floating phase by extracting the central charge using Eq.~\eqref{eq10}. 
According to the CFT, the relation between the entanglement entropy of a subsystem $S_A$ and the subsystem size $L_A$ is given as \cite{EE-CFT1, EE-CFT2, EE-CFT3, EE-CFT4, EE-CFT5, EE-CFT6}
\begin{eqnarray}
    S_A \sim \frac{c}{6}\log_2\left[\frac{2N}{\pi}\sin\left(\frac{\pi L_A}{N}\right)\right] + \log_2 g+\frac{s_1}{2}, \label{eq10}
\end{eqnarray}
where $N$ is the system size and $g, s_1$ are non-universal constants.
The extracted central charge is $c \approx 1.037$, which can be interpreted by approximately regarding the two-species Rydberg atom array as two interacting critical Ising chains \cite{ANNNI-model2, Ising-ladder}. 


With further increased $\Delta / \Omega$, we identify another new phase between the ordered and disordered phases, 
which has never been observed in single-species Rydberg atom array. 
In this phase, the population of the A atoms $\langle \hat{n}_i\rangle_A$ breaks translational symmetry but that of the B atoms is uniform, as demonstrated in Fig.~\ref{phase-diagram}(e).
Since the A sublattice persists the density-wave-like state, the increased excitation density mainly concentrates in the B atoms.
For this reason, we dub this phase as an order-disorder mixture phase.
In such a state, the mirror symmetry with respect to a B atom is simultaneously broken, which can be also characterized by the bond energy of neighboring B atoms. 
Interestingly, we find that althought the local excitation density $\langle \hat{n}_i \rangle$ are invariant for B atoms, the transverse component $\langle\hat{B}\rangle \equiv \langle \hat{\sigma}^+_i \hat{\sigma}^-_{i+2} + h.c. \rangle$ can show a weak oscillation.
As shown in Fig.~\ref{phase-diagram}(e), $\langle \hat{B}\rangle$ of the two neighboring B atoms with an excited A atom in the middle ($\langle \hat{B}\rangle_1$) is slightly lower than that with an A atom possessing very small density ($\langle \hat{B}\rangle_2$), characterizing the broken mirror symmetry with respect to the B atom. 

Another important feature of the phase diagram is the presence of two multicritical points at the boundaries of the order-disorder mixture phase, as shwon in Fig.~\ref{phase-diagram}(d). Notably, the point $P_1$ lies at the intersection of the ordered, order-disorder mixture, and floating phases, while the second multicritical point $P_2$ is enclosed by the floating, order-disorder mixture, and disordered phases. The latter corresponds, in the conventional sense, to a Lifshitz point \cite{Selke1988}. 
Also in Fig.~\ref{phase-diagram}(d), we show the types of different phase boundaries. The disorder to floating transition is
in the Kosterlitz-Thouless (KT) universality class \cite{Kosterlitz1973}. The floating to order and mixture transition is the Pokrovsky-Talapov (PT) transition \cite{Pokrovsky1979}. The order-disorder mixture to order and disorder transition belongs to the common Ising universality class. These results highlight the rich interplay of competing orders in the two-species array and provide a comprehensive understanding of the universality classes governing the phase boundaries.

We also study the phase diagram using mean-field method which 
is able to capture almost all phases but fails to predict the floating phase.
Phase diagram of the Hamiltonian \eqref{Hamiltonian} with 
the $C_6$ coefficients from the second line of Table \ref{interaction-data} is also calculated 
which is similar to the period-4 case. The main difference is the more 
obvious order-disorder mixture (more precisely, half-ordered)
phase area between ordered and disordered phase. 
See Appendix C for more information.

\begin{figure}
    \includegraphics[scale=0.48]{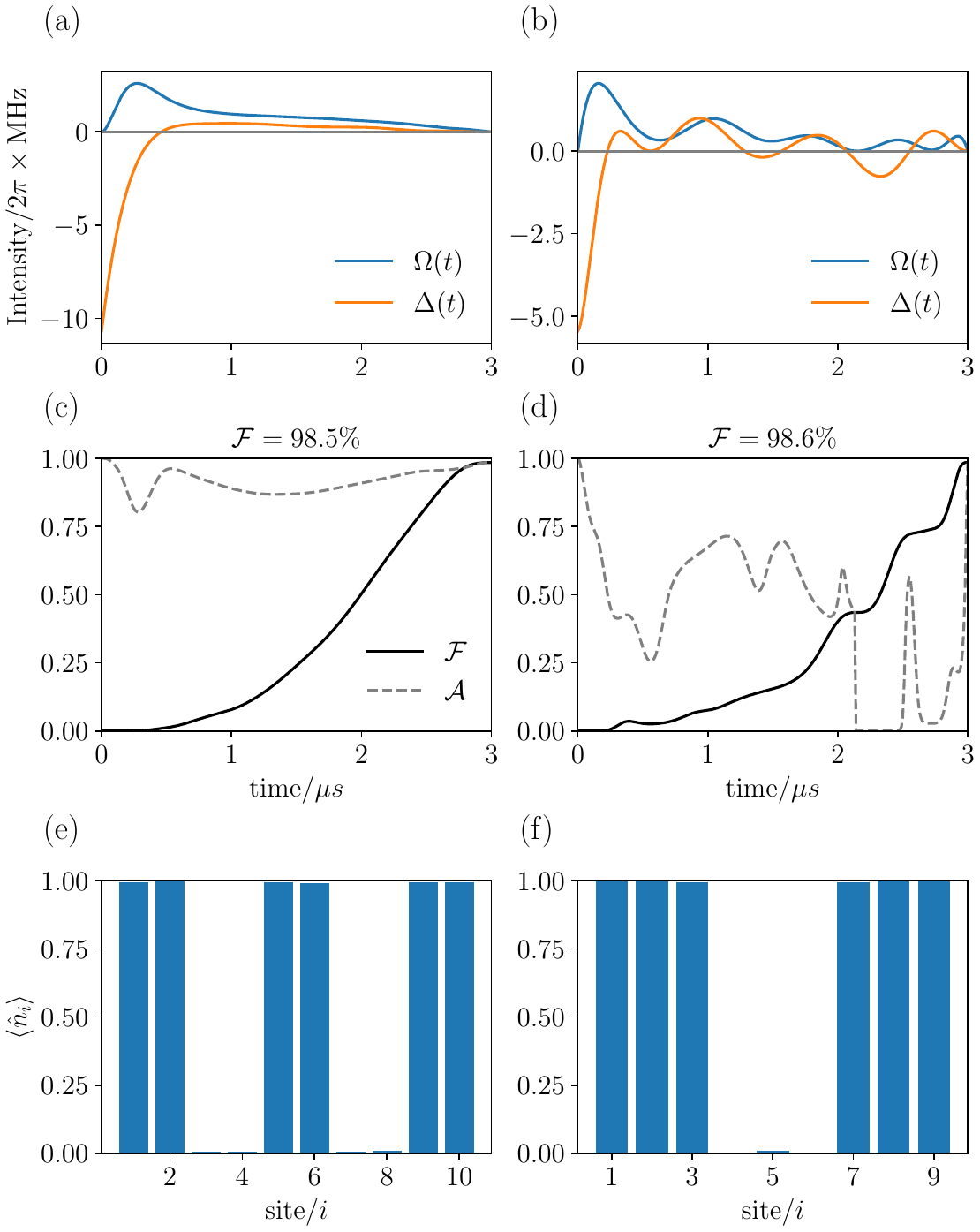}
    \caption{
    The optimization results of $\ket{1100\cdots}$ (left)
    and $\ket{111000\cdots}$ (right) preparation
    in small systems ($N = 10$ for $\ket{1100\cdots}$ and 
    $N = 9$ for $\ket{111000\cdots}$). 
    The first row shows the required pulse of $\Omega(t), \Delta(t)$ and 
    the second row shows the fidelity $\mathcal{F}(t)$ as a function of time. 
    To determine whether the system evolves adiabatically, we also present $\mathcal{A}(t)$, which is defined as the overlap between the instantaneous state and the ground state corresponding to the current parameters. 
    The population $\langle \hat{n}_i\rangle$ 
    of the prepared state is depicted in the last row. 
    The total evolution time $T$ for both 
    states is $T = 3\ \mu s$ and we choose $d=5\ \mu m$ ($d=3\ \mu m$)
    for $\ket{1100\cdots}$ ($\ket{111000\cdots}$) state preparation.
    \label{optimization-result}
    }
\end{figure}

\begin{figure}
    \includegraphics[scale=0.398]{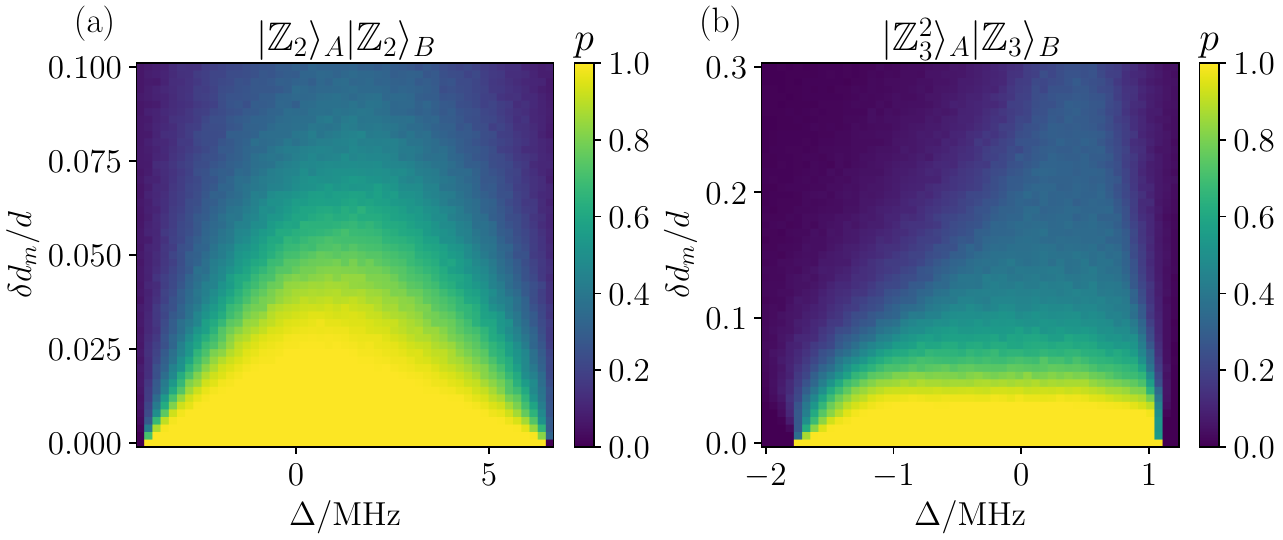}
    \caption{
    Numerical simulations of the robustness of $\ket{1100\cdots}$ (a) 
    and $\ket{111000\cdots}$ (b) ordered states where 
    at each point $(\Delta, \delta d_m/d)$ the real distances between 
    neighbouring atoms $\{d_n\}$ are randomly generated for 2000 times with 
    $|d_n-d|<\delta d_m, \forall n$ and the situations that the 
    system still holds ordered ground states are counted to 
    calculate the survival probability $p\in[0, 1]$ of the ordered 
    ground states depicted above. 
    \label{robustness}}
\end{figure}

\section{Ground State Preparation}
The newly-found novel ground states including $\ket{\mathbb{Z}_2}_A\ket{\mathbb{Z}_2}_B$ and 
$\ket{\mathbb{Z}_3^2}_A\ket{\mathbb{Z}_3}_B$ can be prepared in experiments. 
To be specific, we begin from 
product state $\ket{0000\cdots}$ where there is barely no interaction 
between atoms and turn on the laser to let the system evolve to 
the target states via some controlled time-dependent external fields. 
We define the fidelity function as 
\begin{eqnarray}
    & & \mathcal{F}[\Omega(t), \Delta(t)] = |\langle\psi_{t}|\hat{U}|\psi_0\rangle|^2, \\
    & & \hat{U} = \mathcal{T}\exp\left(-i\int_0^T\mathrm{d}t\ \hat{H}[\Omega(t), \Delta(t)]\right), 
\end{eqnarray}
where $\hat{H}(t)$ is the system Hamiltonian \eqref{Hamiltonian}, $\mathcal{T}$ is the 
time-ordering operator, $T$ is the total evolution time and $\ket{\psi_0}$($\ket{\psi_t}$) 
is the initial(target) state. 
Fidelity, which we aim to maximize, is a functional depending on $\Omega(t)$ and $\Delta(t)$.
In order to find the control pulse $\Omega(t)$ and $\Delta(t)$ for the time evolution required, 
we use dressed Chopped RAndom Basis(dCRAB) method~\cite{dCRAB1, dCRAB2}. 
For states in the ordered phase such as $\ket{1100\cdots}$ and $\ket{111000\cdots}$, using 
dCRAB method, we found some pulses for $\Omega(t)$ and $\Delta(t)$ such that 
the required states can be obtained 
after time evolution. These pulses are shown in Fig.~\ref{optimization-result}.

The robustness of these novel ordered ground states is vital to 
their experimental preparation. We thus study the robustness of these states where 
we mainly consider the imperfection of interatomic spacings. In experiments, the 
atom will deviate from equilibrium position due to thermal fluctuations and the influence from environment 
leading to the slight change in interaction strength. 
The robustness of the period-4 and period-6 ordered states are shown in 
Fig.~\ref{robustness} where the vertical axis $\delta d_m/d$ represents 
the maximum distance of deviation from the central position in both $x$ and $y$ directions. 
The numerical results indicates that both $\ket{1100\cdots}$ and $\ket{111000\cdots}$ 
states are robust in the situation considered.

\section{Quench Dynamincs}
\begin{figure*}
    \includegraphics[scale=0.47]{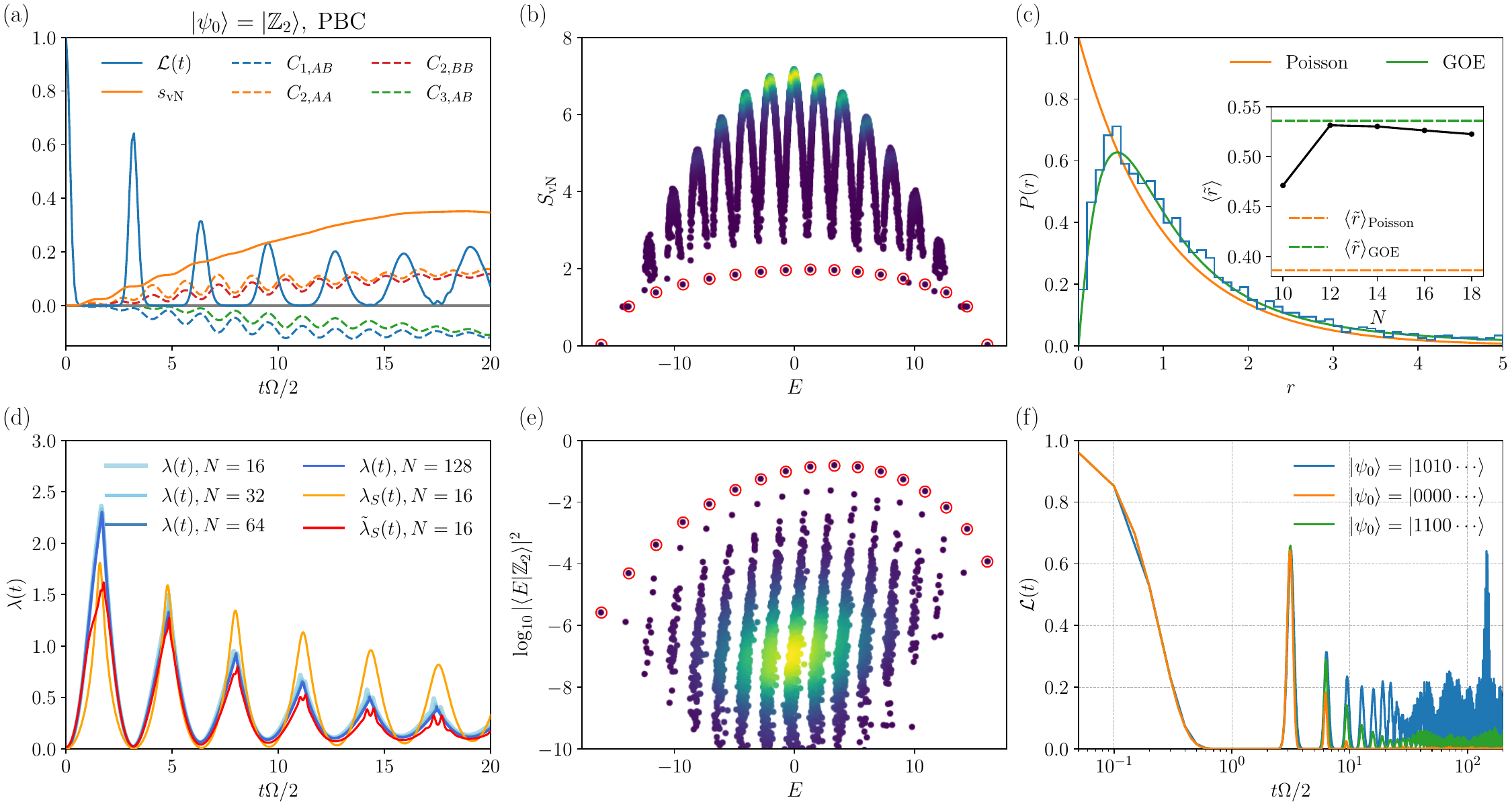}
    \caption{Evidence of the existence of quantum many-body scars 
    in the two-species Rydberg atoms array. We fix atom number $N=16$, 
    atom spacing $d=9\ \mu m$ and use the data 
    from the first row of Table \ref{interaction-data}. The 
    external field strength is $(\Omega, \Delta) = (2, 0.05)\ \mathrm{MHz}$.
    (a) Quantum quench from initial state $\ket{\mathbb{Z}_2}$ 
    demonstrated by Loschmidt echo $\mathcal{L}(t)$, bipatite 
    entanglement entropy per atom  $s_{\mathrm{vN}}$ and four different 
    correlation functions defined as $C_{\delta,ij}=\langle n^i_{0}n^j_{\delta}\rangle - \langle n^i_0\rangle\langle n^j_{\delta}\rangle$ where $i, j$ denotes the corresponding atom species. 
    (b) Bipartite entanglement entropies of all the eigenstates. Those 
    approximate quantum many-body scar states are marked with red circles. 
    The colour scheme illustrates the density of points. A brighter area indicates a higher point density.
    (c) Level spacing distribution of the system in the symmetry reduced sector (momentum $k=0$, inversion even sector).
    Inset: Level spacing indicator $\langle\tilde{r}_i\rangle$ as a function of system size $N$. 
    Orange(Green) dashed line is given by Poisson(GOE) distribution. 
    (d) The Loschmidt rate $\lambda(t)$ exhibits non-analytical behavior for initial 
    N\'eel state showing there is a dynamical quantum phase transition. The approximate Loschmidt rate calculated based on the scarred eigenstates is consistent with the exact one where $\lambda_S(t)$ is the analytical result derived from Eq.~\eqref{eq17} based on $5$ selected scarred eigenstates (see main text and Fig.~\ref{scar_overlap}) while $\tilde{\lambda}_S(r)$ is obtained via numerical method based on all scarred eigenstates. 
    (e) Overlaps between $\ket{\mathbb{Z}_2}$ and scar states are marked with red circles and 
    they contribute more significantly to the system dynamics than other eigenstates(unmarked dots).
    (f) The quantum quench starting from the empty state $\ket{\psi_0} = \ket{0000\cdots}$ exhibits rapid thermalization, whereas the quench from $\ket{1010\cdots}$ and $\ket{1100\cdots}$ does not.
    \label{quantum-quench}
    }
\end{figure*}

It's believed that the time evolution of complex quantum systems is dominated by 
the eigenstates thermalization hypothesis (ETH), i.e., 
systems which are initially prepared in far-from-equilibrium states can evolve 
in time to a state which appears to be in thermal equilibrium and such systems 
are called quantum ergodic \cite{ETH1, ETH2, ETH3}. 
However, not all quantum systems obey ETH. 
The most widely known examples include integrable models and many-body localized systems where 
the symmetry or the conservation law plays an important role against 
ETH thus leading to strong ergodicity breaking \cite{MBL1, QMBS1}.

It has been recently observed that the one dimensional 
Rydberg atom arrays exhibit non-thermalizing phenomenon 
for some initial states such as $\ket{\mathbb{Z}_2}=\ket{1010\cdots}$, 
i.e., the Loschmidt echo shows revivals over time \cite{Lukin201711}. 
There has been plenty of theoretical 
studies of this phenomenon and it's found that in the spectrum of the low energy effective Hamiltonian of 
the Rydberg atom arrays (the ``PXP'' model), there exists some slightly entangled eigenstates that have 
enhanced overlaps with initial state $\ket{\mathbb{Z}_2}$ than other eigenstates 
\cite{Papic201810,Lukin201901,Papic202106,Papic201807}. 
These eigenstates are called quantum many-body scars (QMBS) and they cause weak ergodicity breaking 
in the single-species Rydberg atom arrays. 
In the two-species Rydberg atom arrays, we find a similar phenomenon in the time evolution and 
the approximate QMBS in the system spectrum can be well understood in the mean-field framework 
which is different from the one-species case.

The quench dynamics of the two-species system also exhibits pronounced nonthermal behavior for certain product initial states. Starting from the N\'eel state
$\ket{\psi_0}=\ket{\mathbb{Z}_2}=\ket{1010\cdots}$, 
the Loschmidt echo $\mathcal{L}(t)=|\langle\psi_0|e^{-i\hat{H}t}|\psi_0\rangle|^2$, shows clear revivals during the time evolution, as illustrated in Fig.\ref{quantum-quench}(a). In addition, we find that the product state $\ket{1100\cdots}$ also displays slower thermalization than the trivial vacuum state $\ket{00\cdots}$, see Fig.~\ref{quantum-quench}(f). 
Importantly, this anomalous dynamical behavior does not originate from integrability. As shown in Fig.\ref{quantum-quench}(c), the distribution of the ratio of consecutive level spacings, $P(r)$, is well described by random-matrix-theory statistics rather than a Poisson distribution \cite{level_spacing_indicator}, indicating that the system is nonintegrable. 
The revivals of the Loschmidt echo and the absence of thermalization for specific initial states therefore constitute a genuinely nontrivial phenomenon. 
To uncover its origin, we numerically diagonalize the Hamiltonian and analyze the structure of its eigenstates. As shown in Fig.~\ref{quantum-quench}(b) and (e), we identify a special subset of eigenstates with significantly lower bipartite entanglement entropy than the bulk of the spectrum. These states also exhibit anomalously large overlap with the initial N'eel state $\ket{\mathbb{Z}_2}$. The coexistence of low entanglement and enhanced overlap strongly suggests that these states are approximate quantum many-body scars. Furthermore, they can be approximately understood within a mean-field picture as quasiparticle-like excitations built on top of a special low-entanglement manifold. 

To provide an analytical explanation for these phenomena, we map the Rydberg system Hamiltonian onto a simpler model, the quantum ANNNI model. This mapping is justified, as the long-range interactions have negligible impact on the system dynamics, and the two $C_6$ coefficients, $C^{AA}_6$ and $C^{BB}_6$, are sufficiently close that their difference can be neglected in this situation. 
The quantum ANNNI model Hamiltonian is a Kitaev chain \cite{Kitaev2001,KitaevChain} after performing Jordan-Wigner transformation and the mean-field approximation. 
The simplified effective Hamiltonian in terms of spinless fermion operators 
$\hat{c}_n$ reads
\begin{eqnarray}
    \hat{H} &=& \frac{\tilde{\Omega}}{2}\sum_n
    (2\hat{c}_n^{\dagger}\hat{c}_n-1)+V_1\sum_n(\hat{c}_n-\hat{c}_n^{\dagger})(\hat{c}_{n+1}^{\dagger}+\hat{c}_{n+1}) \nonumber \\
    & &+\tilde{V}_2\sum_n(\hat{c}_n-\hat{c}_n^{\dagger})(\hat{c}_{n+2}^{\dagger}+\hat{c}_{n+2}), 
\end{eqnarray}
where $\tilde{\Omega} = \Omega + 2\mu V_2$, $V_1$, and $\tilde{V}_2 = \alpha V_2$ 
govern the chemical potential, NN and NNN hopping, 
as well as the pair creation and annihilation terms respectively. The mean-field variational 
parameters $\alpha$ and $\mu$ are defined in Appendix F.

We find that in the simplified model the low bipartite entanglement entropy eigenstates 
(denoted $\ket{s_n}, n = 0, 1, \cdots, N$) also exist and they 
mainly fall into the subspace composed of zero-momentum 
Cooper pairs ($\hat{\beta}^{\dagger}_k\hat{\beta}^{\dagger}_{-k}$) and single fermion ($\hat{c}_{k=0}^{\dagger}$), 
i.e., they can be well constructed using the states $\ket{\psi_{\{n_k\}}^{\mathrm{ABC}}}$ for even $n$ 
or $\ket{\psi_{\{n_k\}}^{\mathrm{PBC}}}$ for odd $n$ in Eq.~\eqref{quasiparicle-state}, 
\begin{eqnarray}
    \ket{\psi_{\{n_k\}}^{\mathrm{ABC}/\mathrm{PBC}}} &=& 
    \prod_{k}^{\mathcal{K}^{\mathrm{ABC}/\mathrm{PBC}}}
    (\hat{\beta}_k^{\dagger}\hat{\beta}_{-k}^{\dagger})^{n_k}
    \ket{\emptyset}^{\mathrm{ABC}/\mathrm{PBC}} \label{quasiparicle-state}, 
\end{eqnarray}
where $\mathcal{K}^{\mathrm{ABC}} = \left\{k=(2m-1)\pi/N, m = 1, \cdots, N/2\right\}$, 
$\mathcal{K}^{\mathrm{PBC}} = \left\{k=2m\pi/N, m = 1, \cdots, N/2-1\right\}$, 
$\hat{\beta}_k^{\dagger}\hat{\beta}_{-k}^{\dagger}$ is 
the excitation operator of Cooper pair with momentum $(k, -k)$, $n_k = 0, 1$ is the 
number of Cooper pair excitation at $(k,-k)$, 
$\ket{\emptyset}^{\mathrm{ABC}}$ 
is the ground state of the system in anti-periodic 
boundary condition (ABC) sector and $\ket{\emptyset}^{\mathrm{PBC}}$ 
is the ground state in PBC sector. 
The overlap between the states constructed from quasi-particle operators $\ket{s'_n}$ and 
the exact low-entanglement entropy eigenstates $\ket{s_n}$ obtained via exact diagonalization (ED) 
in the small-size ANNNI model ($N = 16$), as shown in 
Fig.~\ref{ANNNIfig}(d),  is at least 0.89, 
\begin{eqnarray}
    |\braket{s_n|s'_n}|^2 > 0.89  \quad \text{for all } n.
\end{eqnarray}
This strong overlap provides a significant explanation for these approximate quantum many-body scars, 
suggesting that the system predominantly evolves within the zero-momentum symmetry sector when the 
initial state is $\ket{\mathbb{Z}_2}$. 

As the system size increases, the revival of the Loschmidt 
echo $g(t)$ becomes less pronounced. 
However, by defining the Loschmidt rate for finite-size systems \cite{DQPT1}
\begin{equation}
\lambda(t) = -\frac{1}{N} \ln \mathcal{L}(t),
\end{equation}
we observe discontinuities in $\lambda(t)$ over time, as shown in Fig.~\ref{quantum-quench}(d). 
These discontinuities suggest the occurrence of a dynamical quantum phase transition (DQPT) 
at specific time points, which can be explained primarily by the contribution of approximate 
quantum many-body scars (QMBS) to the system’s dynamics. 

The initial state $\ket{\mathbb{Z}_2}$ can be formally decomposed by the eigenstates $\ket{E_k}$ and also can be 
approximately decomposed by the scar states $\ket{E_s}$ where we use $k$ to label general eigenstates and $s$ to label scar eigenstates 
\begin{equation}
    \ket{\mathbb{Z}_2} = \sum_{k=0}^{2^N-1} c_k\ket{E_k}\approx \sum_{s=0}^{N} c_s\ket{E_s}, 
\end{equation}
and we reach the final expression for Loschmidt echo 
\begin{equation}
    \mathcal{L}_S(t)=\sum_{s=7}^{10}2\Big( |c_{s}|^2|c_{s+1}|^2\cos\Delta E_s t + |c_s|^4
    \Big), \label{eq17}
\end{equation}
by only considering contributions from the five scar eigenstates that have the largest overlap with $\ket{\mathbb{Z}_2}$ (for a $N = 16$ system under PBC), and neglecting the high-frequency components of the dynamics. 
After proper normalization, the Loschmidt rate $\lambda_S(t)$ calculated from Eq.~\eqref{eq17} are 
highly consistent with the exact one as shwon in Fig.~\ref{quantum-quench}(d). Further discussion on the relationship between QMBS and DQPT can be found in Appendix H.

\section{Applications in combinatorial optimization problems}

The search for the ground state of an interacting many-body system is closely related to solving discrete optimization problems. In programmable atomic arrays, the energy of a given atomic configuration can be interpreted as the cost function of a classical optimization task, such that the ground state corresponds to the optimal solution. A prominent example is the mapping between single-species Rydberg atom arrays and the maximum independent set (MIS) problem on a graph \cite{Ebadi2022,Byun2022}. In this case, the Rydberg blockade mechanism prevents nearby atoms from being simultaneously excited, which directly enforces the independent-set constraint on vertices of the graph.

The introduction of multiple atomic species enriches the interaction structure of the system and therefore enlarges the class of optimization problems that can be encoded. Unlike single-species Rydberg arrays, where the interactions between Rydberg excitations are purely repulsive, two-species systems can exhibit both effective repulsive and attractive interactions depending on the species involved. This additional flexibility allows the effective energy function of the atomic array to realize a more general quadratic binary optimization problem.

To illustrate this mapping, we consider the classical limit of the two-species Rydberg Hamiltonian
\begin{equation}
H = \sum_i \delta_in_i+\sum_{i<j} V_{ij} n_i n_j ,
\end{equation}
where $n_i\in\{0,1\}$ denotes the excitation number of atom at site $i$, $\delta_i$ is the local detuning and $V_{ij}$ describes the interaction between atoms of different species. Because the interaction coefficients $V_{ij}$ can take different magnitudes and signs depending on the species combination, the resulting classical energy landscape can contain both positive and negative couplings.

By introducing Ising variables $s_i = 2n_i - 1 \in \{-1,1\}$,
the Hamiltonian can be rewritten in the quadratic Ising form
\begin{equation}
H_{\mathrm{Ising}} = \sum_i h_i s_i + \sum_{i<j} J_{ij} s_i s_j ,
\end{equation}
which is equivalent to the standard quadratic unconstrained binary optimization (QUBO) formulation widely used in combinatorial optimization \cite{Fred2019}. Different choices of the effective couplings $J_{ij}$ therefore correspond to different optimization problems.

A particularly natural example is the signed graph partition problem. Consider a graph $G=(V,E)$ whose edges carry positive or negative weights. The goal is to partition the vertices into two groups such that positively weighted edges favor vertices belonging to the same group, while negatively weighted edges favor vertices belonging to different groups. Using Ising variables $s_i \in \{-1,1\}$ 
to label the two partitions, one may write the corresponding cost function as
\begin{equation}
C_{\mathrm{SGP}}(\{s_i\}) = - \sum_{(i,j)\in E} J_{ij} s_i s_j ,
\label{eq:sgp}
\end{equation}
where $J_{ij}$ can be either positive or negative. If $J_{ij}>0$, the cost is minimized by $s_i=s_j$, so the two vertices prefer to lie in the same partition. If $J_{ij}<0$, the cost is minimized by $s_i=-s_j$, so the two vertices prefer to lie in different partitions. In this way, the ground state of Eq.~(\ref{eq:sgp}) directly gives the optimal signed partition of the graph.

\begin{figure}
    \includegraphics[scale=1]{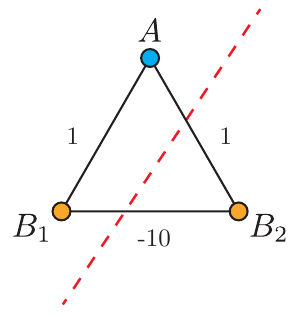}
    \caption{
    Illustration of a simple signed graph partition realized in a two-species 
    Rydberg array. Attractive interactions between the A atom and the two B atoms 
    are represented by positive edges, while the B-B interaction is repulsive. 
    The ground-state configuration corresponds to exciting one A atom and one B atom, 
    which gives the optimal partition of the graph.
    \label{sgp}}
\end{figure}

We can consider a triangular cluster consisting of one atom of species A and two atoms of species B, as shown in Fig.~\ref{sgp}. Suppose that the interaction between A and B atoms is attractive, while the interaction between the two B atoms is repulsive. In graph language, this corresponds to a triangle with two positive edges (between A and each B) and one negative edge (between the two B atoms). The corresponding cost function may be written as
\begin{equation}
C = -J_{AB}(s_A s_{B_1} + s_A s_{B_2}) - J_{BB} s_{B_1}s_{B_2},
\end{equation}
with $J_{AB}>0$ and $J_{BB}<0$. The positive couplings $J_{AB}$ favor A and each B being in the same partition, whereas the negative coupling $J_{BB}$ favors the two B atoms being in opposite partitions. These conditions cannot all be satisfied simultaneously, so the system is frustrated. The optimal configuration is obtained by placing A and one of the B atoms in the same partition, while the other B atom belongs to the opposite partition.

The same structure can be understood directly in the Rydberg language. Because the A-B interaction is attractive, the energy is lowered when one A atom and one B atom are simultaneously excited. On the other hand, because the B-B interaction is repulsive, the energy is increased if both B atoms are excited at the same time. As a result, the lowest-energy configuration is the one in which one A atom and one B atom are excited, while the other B atom remains unexcited. This realizes, in the simplest possible setting, the solution of a signed graph partition problem through the ground-state configuration of a two-species Rydberg array.

Therefore, in contrast to single-species Rydberg arrays, which mainly implement hard-constraint optimization problems such as MIS, two-species Rydberg systems can naturally encode optimization problems with both positive and negative couplings. This substantially broadens the class of combinatorial optimization problems that can be addressed using programmable neutral-atom platforms.

\section{Discussion and Outlook}
We investigate the phase diagram of a two-species Rydberg atom array and 
discover a variety of novel ground state configurations that do not occur 
in single-species arrays, 
such as the period-4 and period-6 product states. Through adiabatic evolution, 
assisted by optimal control techniques, we propose implementable control schemes 
to prepare those novel ordered states. 
Additionally, we test the robustness of these states against 
atomic deviation, which is the most critical imperfection in Rydberg atom arrays. 
The interplay between quantum fluctuations and atomic interactions also leads to other emergent phases, 
including a unique order-disorder mixture phase and a floating phase. 
The former has not been previously reported in cold atom systems or other quantum simulation platforms. 
While the floating phase in single-species arrays has been extensively studied 
and was recently observed in ladder atom arrays \cite{Jin2024}, we note that, in contrast to the single-species 
case where the floating phase is described by a free bosonic field theory, the two-species system 
requires additional interaction terms. Although the low-energy effective theory of the two-species system 
is still dominated by a free bosonic field, the renormalization analysis reveals richer theoretical 
insights into the structure of the entire phase diagram.

The two-species Rydberg atom array can also exhibit quantum many-body scars 
in the energy spectrum similar to the PXP model 
and their large overlap with initial $\ket{\mathbb{Z}_2}$=$\ket{1010\cdots}$ state 
can also induce revivals in Loschmidt echo over time. 
We approximately map two-species atom array Hamiltonian to 
the quantum ANNNI model Hamiltonian and found that the time evolution of $\ket{\mathbb{Z}_2}$ 
is mainly restricted in the momentum-zero subspace and the scarred eigenstates can be well 
constructed using the quasi-particle operators of Cooper pair form with zero momentum. 
We stress that unlike ``PXP'' model the ergodicity breaking phenomenon here is induced by the proper 
perturbations instead of the Rydberg blockade mechanism. Furthermore, it should also be 
emphasized that the DQPT found in the same system can be explained by the newly discovered 
scarred states. This connection between DQPT and QMBS has been pointed out in \cite{Van2023}. 

Our current work lays the foundation for quantum simulation using two-species Rydberg atom array. 
The various novel classical ground state configurations are ideal platform for simulating 
abundant charge density wave states in condensed matter system and our results considering 
quantum fluctuations have also paved the way for studying quantum phase transitions and other critical phenomenon 
in the two-species Rydberg atom array.  Our study of quench dynamics also shows that 
the two-species Rydberg atom array can be utilized to explore the mechanism of ergodicity breaking as well as 
its relation with DQPT. In addition, the richer interaction structure arising from multiple atomic species may also enable the encoding of more general combinatorial optimization problems. Such systems allow the ground-state search to be naturally mapped to tasks such as signed graph partition problems.

In this work we focus primarily on scenarios where all $C_6$ coefficients are effectively comparable, which is crucial for stabilizing the novel phases discussed above. It would be interesting to extend this investigation to cases with different ratios of $C_6$ coefficients. Moreover, exploring different spatial arrangements of multiple atomic species in one-dimensional systems, as well as more complex geometries in two dimensions, may lead to an even richer landscape of quantum phases and dynamical phenomena. The role of dissipation in such systems also remains an important direction for future study.

In conclusion, the two-species Rydberg atom array provides a new platform for exploring complex quantum many-body phenomena. Our work suggests that these phenomena can be convincingly realized in experiments and may also find applications in quantum computation for solving more general combinatorial optimization problems. 

\begin{acknowledgments}
We thank Kun-Peng Wang, Li-Ping Yang and Gao-Yong Sun for useful discussions. 
This work was supported by the National Key Research and Development Program of China (Grant Nos.~2021YFA1402001 and 2021YFA0718304), 
the National Natural Science Foundation of China (NSFC) (Grant Nos.~12375007, 12135018,  12047503, 12074391, 12261131507 and U20A2074), 
CAS Project for Young Scientists in Basic Research (Grant No.~YSBR-057), 
the Special Project in Key Areas for Universities in Guangdong Province (No.~2023ZDZX3054), 
the Major Program (JD) of Hubei Province (Grant No.~2023BAA020) 
and by the 
Dongguan Key Laboratory of Artificial Intelligence Design for Advanced Materials. 

\end{acknowledgments}


\appendix
\section{Discussion on the classical phase diagram}
We consider the classical situation where there is no transverse field $\Omega = 0$.
In order to simplify our discussion, we also set $\Delta = 0$, i.e., we only study the phase diagram of the Hamiltonian $\hat{H}_1$ in Eq.~\eqref{H1}. 
We define $V_{AA} = C_6^{AA}/d^6, V_{BB}=C_6^{BB}/d^6, V_{AB}=C_6^{AB}/d^6$ and choose $V_{AB}$ as the energy unit.
We set $V_{AB} = -1$ to introduce competition of different states. 
The classical phase diagram has already been shown in Fig.~\ref{classical_PD}, where we use excitation density $n\equiv\frac{1}{N}\sum_i\langle \hat{n}_i\rangle$ as the order parameter to distinguish different phases ($N$ is the system size).
In Fig.~\ref{classical_PD}, the system size can be considered as towards infinity $N\to \infty$. 

\begin{figure}
    \centering
    \includegraphics[scale=0.4]{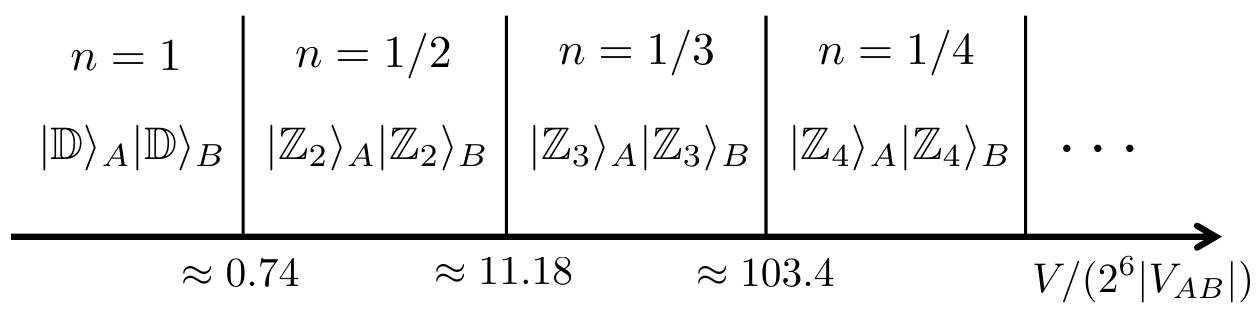}
    \caption{Phase diagram of Hamiltonian $\hat{H}_1$ in Eq.~\eqref{H1} along the white dashed line 
    $V_{AA}=V_{BB}=V$ in Fig.~\ref{classical_PD} with $V_{AB}=C_6^{AB}/d^6=-1$ the energy unit.
    \label{classicalPD_line}}
\end{figure}

The classical phase diagram is symmetric about the line $V_{AA}=V_{BB}$ marked with a white dashed line in Fig.~\ref{classical_PD}. 
We discuss a simpler case with $V_{AA}=V_{BB}=V$. 
Since $V_{AB}=-1<0$, the system will prefer the neighboring excitation, and the positive NNN interaction $V>0$ will lead to the crystalline ground states in each sublattice.  
Therefore, when $V$ is small the system prefers a disordered phase in the sublattices of both A and B atoms denoted as $\ket{\mathbb{D}}_A\ket{\mathbb{D}}_B$, and with growing $V$, the blockade mechanism begins to take effect, giving birth to different crystalline ground states with different excitation density including the abovementioned $\ket{11001100\cdots}$ configuration denoted as $\ket{\mathbb{Z}_2}_A\ket{\mathbb{Z}_2}_B$. 
The phase diagram along the $V_{AA}=V_{BB}$ line is 
summarized in Fig.~\ref{classicalPD_line}. 
When $V_{AA}>0$ and $V_{BB}<0$ (or $V_{AA}<0$ and $V_{BB}>0$), the sublattice of atom A still tends to stay in crystalline states but the B atoms prefer a disordered state, which leads to the formation of a ``half-ordered'' phase denoted as $\ket{\mathbb{Z}_q}_A\ket{\mathbb{D}}_B$ (and $\ket{\mathbb{D}}_A\ket{\mathbb{Z}_q}_B$, $q=2, 3,\cdots$).

Numerical simulations uncover two new phases beyond our simple analysis. 
We denote the first new phase as $\ket{\mathbb{Z}_3}_A\ket{\mathbb{Z}_3^2}_B$ (and also $\ket{\mathbb{Z}_3^2}_A\ket{\mathbb{Z}_3}_B$ because of symmetry), where $\ket{\mathbb{Z}_q^p}$ stands for a state with unit cell length $q$ and excitation number $p$.
We use $\ket{\mathbb{Z}_q} = \ket{\mathbb{Z}_q^1}$ for simplicity. 
The $\ket{\mathbb{Z}_3}_A\ket{\mathbb{Z}_3^2}_B$ phase is found at the $V_{AA} \gg V_{BB} > 0$ area, where the interaction strength between A atoms $V_{AA}$ is large enough to maintain a crystalline state $\ket{\mathbb{Z}_3}$ while $V_{BB}$ is too weak to do so, thus the influence of $V_{AB}$ persuades the system to realize a novel $\ket{111000111000\cdots}$ configuration (the first site is B atom here). 
Notice that because of the same excitation density of $\ket{\mathbb{Z}_2}_A\ket{\mathbb{Z}_2}_B$ and $\ket{\mathbb{Z}_3}_A\ket{\mathbb{Z}_3^2}_B$, these two phases share the same color in Fig.~\ref{classical_PD}. 
While the $\ket{\mathbb{Z}_3}_A\ket{\mathbb{Z}_3^2}_B$ state mainly lives around $V_{AA}=0$ and $V_{BB}=0$ lines marked with grey dashed lines, the $\ket{\mathbb{Z}_2}_A\ket{\mathbb{Z}_2}_B$ state mainly dominates the area around the white dashed line not including the area close to grey dashed line. 
The second new phase is denoted as $\ket{\mathbb{Z}_3^2}_A\ket{\mathbb{D}}_B$ (and $\ket{\mathbb{D}}_A\ket{\mathbb{Z}_3^2}_B$), and it is located at the boundary between $\ket{\mathbb{D}}_A\ket{\mathbb{D}}_B$ and $\ket{\mathbb{Z}_2}_A\ket{\mathbb{D}}_B$ 
(also $\ket{\mathbb{D}}_A\ket{\mathbb{Z}_2}_B$). 
Due to the very small area, we do not label this phase 
in Fig.~\ref{classical_PD}.

We do not consider the influence of detuning $\Delta$, but it is worth being pointed out that the interplay between $\Delta$ and $V_{AA},V_{BB},V_{AB}$ will give birth to more complex configurations. 
For example, we find that when $V_{AA}=50\times 2^6, V_{BB}=10\times 2^6, V_{AB}=-1$ and $\Delta = 1$, the ground state appears to be $\ket{\mathbb{Z}_3}_A\ket{\mathbb{Z}_2}_B$ with the excitation density $n=5/12$, which is not found in our analyses at $\Delta=0$.
We leave the work of completing the classical phase diagram to future study, and the current results already suffice to instruct us to find the proper Rydberg states that can produce novel ground state configurations such as the anti-phase $\ket{1100\cdots}$ and etc.

\section{Interaction data for experimental setup}

\begin{figure}[h]
    \centering
    \includegraphics[scale=0.39]{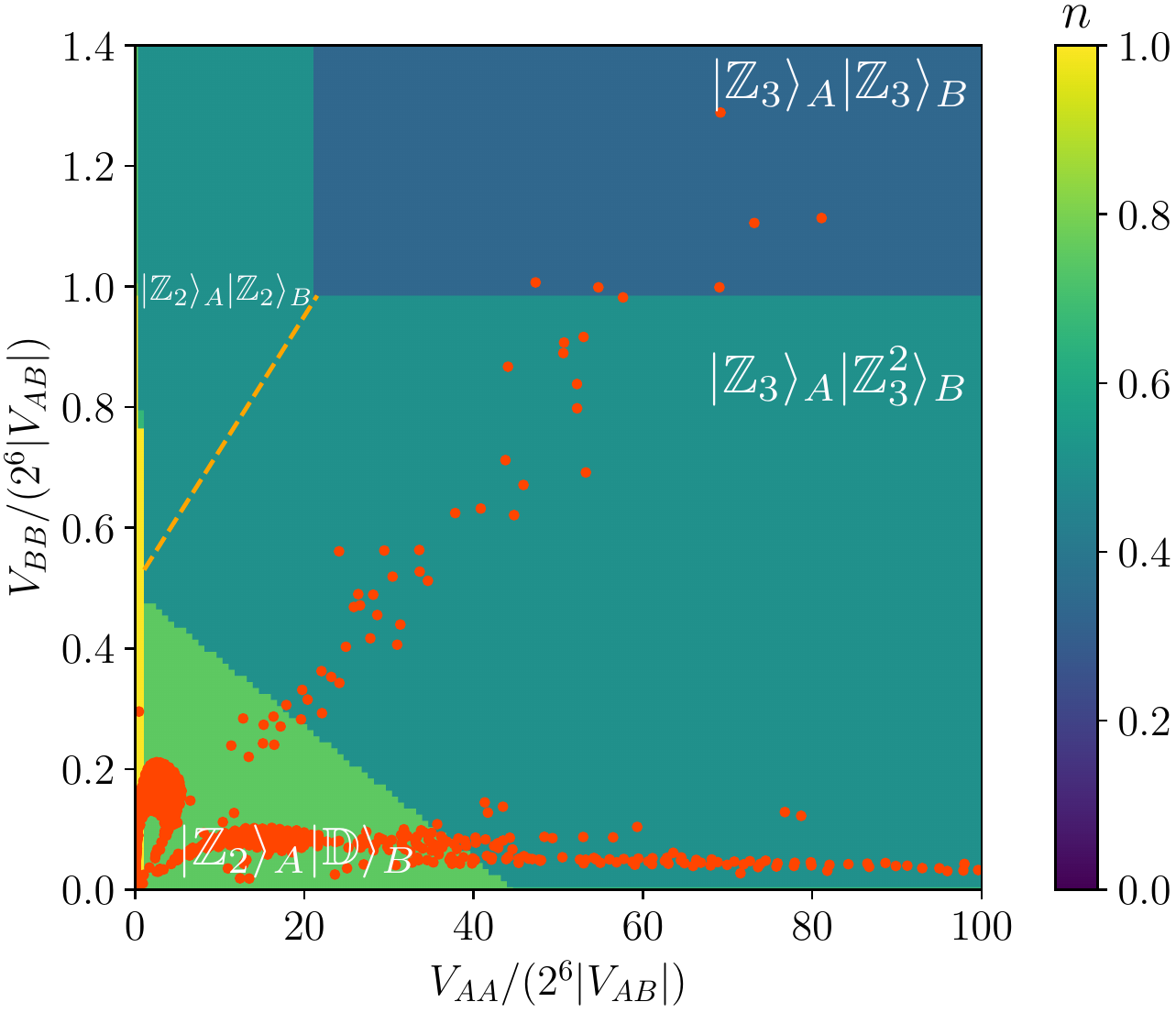}
    \caption{The real atom interaction $C_6$ coefficents are depecited above where $^{85}\mathrm{Rb}$ are denoted 
    A atom and $^{87}\mathrm{Rb}$ B atom here. The orange dashed line marks the boundary separating two distinct phases with the same excitation density. We let $V_{AA}=C_6^{AA}$, $V_{BB}=C_6^{BB}$ and 
    $V_{AB}=C_6^{AB}$ for simplicity and calculate the data points here. 
    \label{cpd_rbrb}}
\end{figure}

\begin{figure}
    \centering
    \includegraphics[scale=0.39]{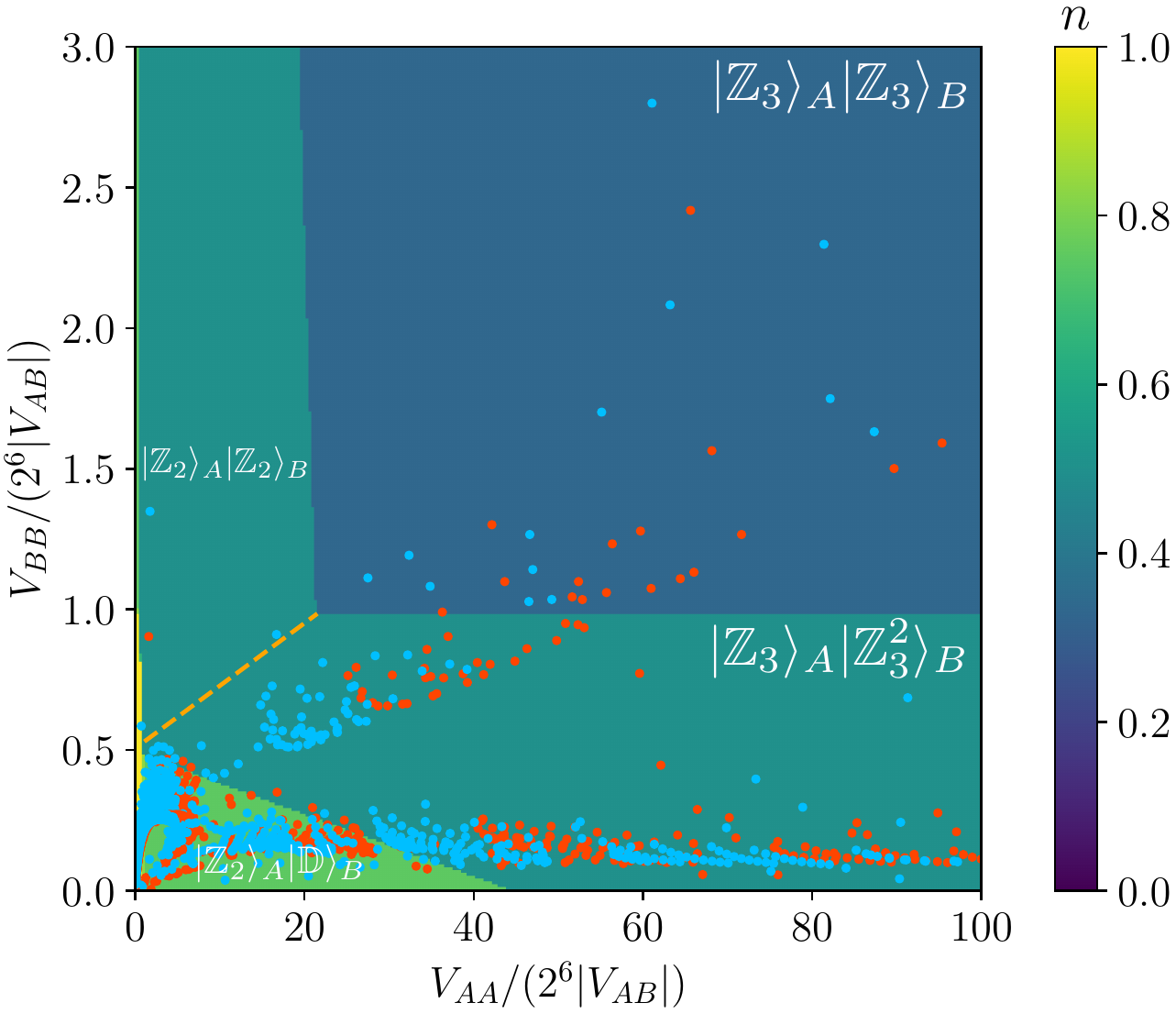}
    \caption{The real atom interaction $C_6$ coefficents are depecited 
    above where $\mathrm{Rb}$ are denoted 
    A atom and $\mathrm{Cs}$ B atom for blue points. For red points, 
    $\mathrm{Cs}$ is the A atom and $\mathrm{Rb}$ is the B atom. 
    The orange dashed line marks the boundary separating two distinct 
    phases with the same excitation density.
    We let $V_{AA}=C_6^{AA}$, $V_{BB}=C_6^{BB}$ and 
    $V_{AB}=C_6^{AB}$ for simplicity and calculate the data points here. 
    \label{cpd_rbcs}}
\end{figure}

Our next step is to check whether it is possible to prepare these states in experiment.
We calculate all the $C_6$ coefficents between the Rydberg states $\ket{n_1 S_{1/2}}_{^{85}\mathrm{Rb}}$ and $\ket{n_2 S_{1/2}}_{^{87}\mathrm{Rb}}$, as well as between $\ket{n_1 S_{1/2}}_{\mathrm{Rb}}$ and $\ket{n_2 S_{1/2}}_{\mathrm{Cs}}$, using the python package ARC \cite{ARC-python-package}, where $n_1, n_2$ stand for the principal quantum number of the first and second atoms ranging from 40 to 90 with $m_j=1/2$ for all the states.
We present the data points giving $C^{AB}_6 < 0$ on the phase diagram. 
The results for $^{85}\mathrm{Rb}$ and $^{87}\mathrm{Rb}$ atoms are shown in Fig.~\ref{cpd_rbrb}, and the results for $\mathrm{Rb}$ and $\mathrm{Cs}$ atoms are shown in Fig.~\ref{cpd_rbcs}. 
Notice that we only show the data points that satisfy $V_{AA}/(2^6|V_{AB}|) < 100$, and we note that there are many data points are beyond this area but our figures are able to cover all the situations.

It can be concluded that for the atoms of the same element (for Rubidium atoms at least), it is almost impossible to prepare a $\ket{\mathbb{Z}_2}_A\ket{\mathbb{Z}_2}_B$ state only using the $S$ states. 
Because if the interaction between two $\mathrm{Rb}$ atoms at the $S$ state is strong ($V_{AA}, V_{BB} > 0 $ are very large), the interaction $V_{AB}$ tends be positive, leading to the empty ground state $\ket{0000\cdots}$ instead of $\ket{1100\cdots}$ state. 
To ensure $V_{AB}<0$ is always maintained, the two principal quantum numbers have to satisfy $|n_1-n_2|\gg 0$, and thus there will always be a small $V_{BB}$ and a large $V_{AA}$, which can only gives birth to other ground state configurations as shown in Fig.~\ref{cpd_rbrb}.

For $\mathrm{Rb}$ and $\mathrm{Cs}$ atoms, the above analysis also applies but numerical simulation finds some $C_6$ coefficents that can produce the  $\ket{\mathbb{Z}_2}_A\ket{\mathbb{Z}_2}_B$ ground state, which indicates it is possible to prepare this anti-phase in experiments.

\section{Quantum phase diagram of the second parameter setup}
\begin{figure}
    \includegraphics[scale=0.88]{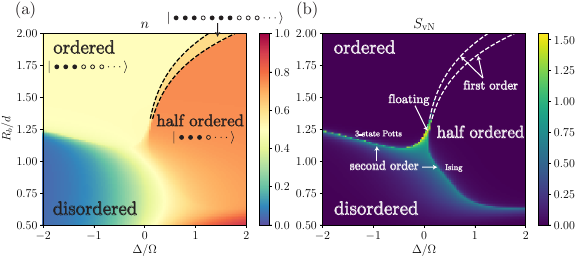}
    \caption{
    DMRG phase diagram of the Hamiltonian \eqref{Hamiltonian} with the
    $C_6$ coefficients from the second line of 
    Table \ref{interaction-data}. We choose the 
    system size to be $N = 267$ which can accommodate an integer number of different unit cells, i.e., 
    $N = 267$ can be decomposed into the form of $6n+3$, $10m+7$ and $4k+3$ with $n, m, k\in\mathbb{N}^+$. 
    (a) First order transition lines marked with two dashed lines can be easily identified using the excitation density $n$. 
    (b) Second order phase transition lines are captured by the bipartite entanglement entropy $S_{\mathrm{vN}}$. 
    \label{phase-diagram-2}
    }
\end{figure}

\begin{figure}
    \includegraphics[scale=0.93]{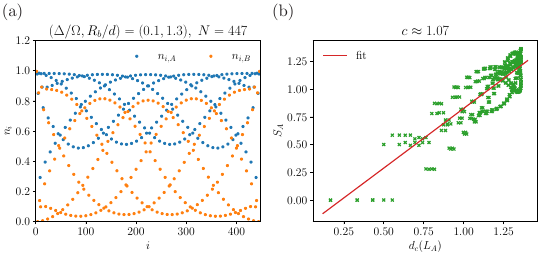}
    \caption{
    State configuration at $(\Delta/\Omega,R_b/d)=(0.1, 1.3)$ in Fig.~\ref{phase-diagram-2} with size $N=447$ is calculated proving the exsistance of the floating phase. 
    (a) The friedel oscillation of local excitation density can be clearly identified. 
    (b) The central charge fitted via Eq.~\eqref{eq10} is $c\approx1.07$ further proving the state here is critical and thus floating. 
    \label{config111000-floating}
    }
\end{figure}

In this section, we discuss the phase diagram of the Hamiltonian defined by the parameters in the second line of Table~\ref{interaction-data}, which can realize the period-6 ground state.
The DMRG results of the excitation density $n$ and entanglement entropy are demonstrated in Fig.~\ref{phase-diagram-2}.
In the orange region of Fig.~\ref{phase-diagram-2}(a), the much larger $C_6$ coefficient between type B atoms leads to the formation of $\ket{\mathbb{Z}_2}$ Rydberg crystal in the sublattice of type B atoms.
And the type A atoms form a full filling disordered state $\ket{1111\cdots}$, giving the total filling factor $3/4$.
Thus, we denote this phase as the half-ordered phase because only half of the system is ordered (B sublattice). 
We also note that there are two different ordered phases. The period-6 ground state 
$\ket{\mathbb{Z}_3^2}_A\ket{\mathbb{Z}_3}_B$ is favored at negative and small positive detunings. 
As the detuning gradually increases, the competition between $\ket{111000\cdots}$ and $\ket{1110\cdots}$ configurations 
leads to a new ordered ground state denoted $\ket{\mathbb{Z}_5^4}_A\ket{\mathbb{Z}_5^2}_B$ with unit cell size 10 
and excitation density 3/5. The phase transitions between ordered phases and half ordered phase 
are all of first order as marked in Fig.~\ref{phase-diagram-2} with dashed lines. 
The floating phase here connects the ordered $\ket{\mathbb{Z}_5^4}_A\ket{\mathbb{Z}_5^2}_B$ phase and the disordered phase as illustrated in Fig.~\ref{config111000-floating}.

\section{Ground-state phase diagram based on the mean-field theory}

In this section, we discuss the mean-field calculate of the phase diagram.
Given the DMRG results, it is natural to choose the unit cell of four lattice spacings in the mean-field framework, i.e., four variational parameters in the following analysis, which is denoted as $\vec{n} = (n_0, n_1,\cdots, n_{L-1})$.

Following the mean-field approximation ($\delta\neq 0$)
\begin{equation}
    \hat{n}_{i}\hat{n}_{i+\delta} \approx \langle\hat{n}_i\rangle \hat{n}_{i+\delta}+
    \hat{n}_{i}\langle \hat{n}_{i+\delta}\rangle - \langle \hat{n}_i\rangle\langle \hat{n}_{i+\delta}\rangle,
\end{equation}
one can obtain the mean-field Hamiltonian for a unit cell
\begin{equation}
    \hat{H}_{\mathrm{MF}} = \hat{H}_0+\hat{H}_1, 
\end{equation}
with the part inside unit cell of length $L$
\begin{equation}
    \hat{H}_0 = \frac{\Omega}{2}\sum_{i=0}^{L-1}\hat{\sigma}_i^x-
    \Delta\sum_{i=0}^{L-1}\hat{n}_i+\sum_{i=0}^{L-2}\sum_{j=i+1}^{L-1}\frac{C_6^{ij}}{d^6|i-j|^6}\hat{n}_i\hat{n}_j,
\end{equation}
and the part derived from mean-field approximation
\begin{eqnarray}
    \hat{H}_1 &=& \sum_{i=0}^{L-1}\sum_{j = 0}^{L-1}
    \frac{C_6^{ij}}{d^6}c_{ij}n_j\hat{n}_i, \\
    c_{ij} &=& \sum_{n = 0}^{+\infty}\left(
    \frac{1}{(6n+L+i-j)^6}+\frac{1}{(6n+L-i+j)^6}
    \right)\nonumber\\
    &=& \frac{1}{120L^6}\left[\psi^{(5)}\left(1+\frac{i-j}{L}\right)+
    \psi^{(5)}\left(1-\frac{i-j}{L}\right)\right]\nonumber, 
\end{eqnarray}
where $C_6^{ij}$ is the $C_6$ coefficient between atom $i$ and $j$, $d$ is 
the atom spacing and $\psi^{(5)}(x)$ is 
the polygamma function of order 5. 

\begin{figure}
    \includegraphics[scale=0.83]{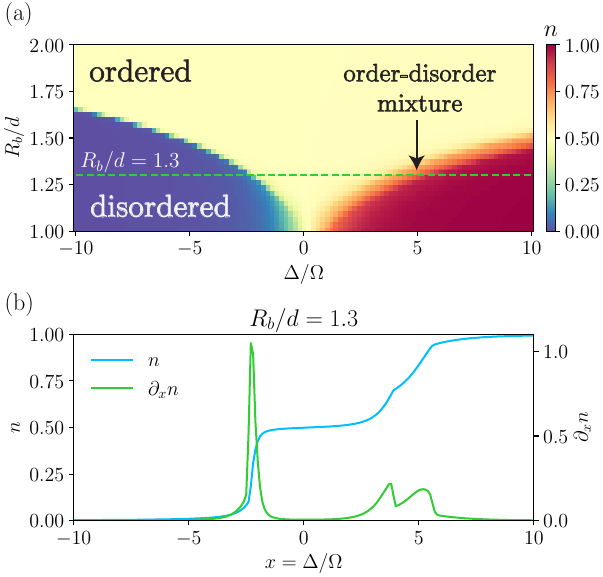}
    \caption{
    Mean field phase diagram of Hamiltonian \eqref{Hamiltonian} with 
    $C_6$ coefficients from the first line of Table \ref{interaction-data}. 
    (a) The excitation density $n$ depecited here is calculated using the optimized 
    mean-field variational parameters $n = \frac{1}{L}\sum_{i=0}^{L-1}n_i$. 
    (b) We also calculate $n$ and its derivative $\partial n/\partial x$ along the 
    green dashed line ($R_b/d=1.3$) in (a) and $x\equiv\Delta/\Omega$ here.
    \label{mean-field-phase-4}
    }
\end{figure}

For $L = 4$, we numerically calculate the optimal choice of variational parameters $\vec{n}$ for the mean-field Hamiltonian $\hat{H}_{\mathrm{MF}}$.
We denote the ground state of $\hat{H}_{\mathrm{MF}}$ as $\ket{\psi_g(\vec{n})}$ and define the 
energy density $\epsilon(\vec{n})$ as well as the fidelity $\mathcal{F}(\vec{n})$ as
\begin{eqnarray}
    \epsilon(\vec{n}) &=& \langle\psi_g(\vec{n})|\hat{H}|\psi_g(\vec{n})\rangle, \\
    1-\mathcal{F}(\vec{n}) &=& \frac{1}{L}\sum_{i=0}^{L-1}(\langle\psi_g(\vec{n})|\hat{n}_i|\psi_g(\vec{n})\rangle-n_i)^2,
\end{eqnarray}
where $\hat{H}$ is the original Hamiltonian of four sites. 
The natural way to find the optimal $\vec{n}$ is to minimize $\epsilon(\vec{n})$. 
However, we found this method is numerical unfriendly, because the optimization algorithms are very likely to be trapped in local minimum and the convergence speed is not very fast due to the possible barren plateaus. 
We use a different method by first minimizing the infidelity $1-\mathcal{F}(\vec{n})$ with many different initial guesses and choosing the one with the lowest energy among these optimization results. 
The obtained phase diagram is shown in Fig.~\ref{mean-field-phase-4}.
One can find that the mean-field theory is able to capture the general picture of the phase diagram and only the floating phase is unable to be explained by this mean-field theory.

\begin{figure}
    \includegraphics[scale=0.39]{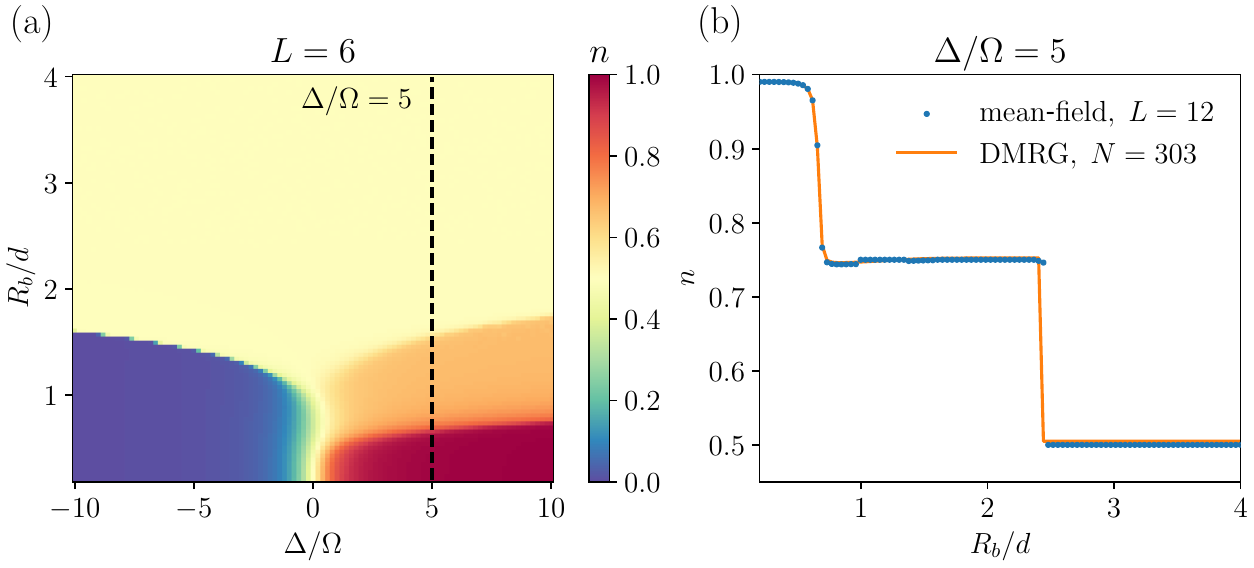}
    \caption{
    (a) Mean field phase diagram of Hamiltonian \eqref{Hamiltonian} with 
    $C_6$ coefficients from the second line of Table \ref{interaction-data}. The order parameter 
    depecited here is calculated using the optimized 
    mean-field variational parameters $\langle n_i\rangle = \frac{1}{L}\sum_{i=0}^{L-1}n_i$.
    (b) The order parameter $\langle n_i\rangle$ calculated using mean-field theory with 
    unit cell size $L = 12$ along the black dashed line ($\Delta/\Omega = 5$) is consistent with
    the DMRG calculations.
    \label{MF-DMRG-6}
    }
\end{figure}

For the paramters with the $\ket{111000\cdots}$ phase, it is most proper to choose $L = 60$, the least common multiple of $4$, $6$ 
and $10$ which can faithfully represent all possible configurations in this situation. 
Since searching the ground state of the Hamiltonian matrix with a $L=60$ unit cell is time consuming, we compromise and calculate the phase diagram using $L = 6$, as shown in Fig.~\ref{MF-DMRG-6}(a), where the boundary of the half-filling phase ($\ket{111000\cdots}$) and $3/4$-filling phase ($\ket{1110\cdots}$) is inconsistent with the DMRG result and the $\ket{\mathbb{Z}_5^4}_A\ket{\mathbb{Z}_5^2}_B$ 
phase are not indicated completely. 
We note that this disagreement is caused by the improper choice of the unit cell size.
For example, by taking $L = 12$, we reexamine one case by scanning $R_b / d$ at fixed $\Delta / \Omega = 5$, it is quite clear that once the unit cell size is properly chosen, the mean-field theory result is highly consistent with that from DMRG, as seen by the comparison in Fig.~\ref{MF-DMRG-6}(b) and it 
is also apparent that the $\ket{\mathbb{Z}_5^4}_A\ket{\mathbb{Z}_5^2}_B$ are not captured by both mean-field 
and DMRG calculations which is still due to the improper choice of the unit cell and system size. 

\section{Finite-size scaling analysis}
\begin{figure}
    \includegraphics[scale=0.49]{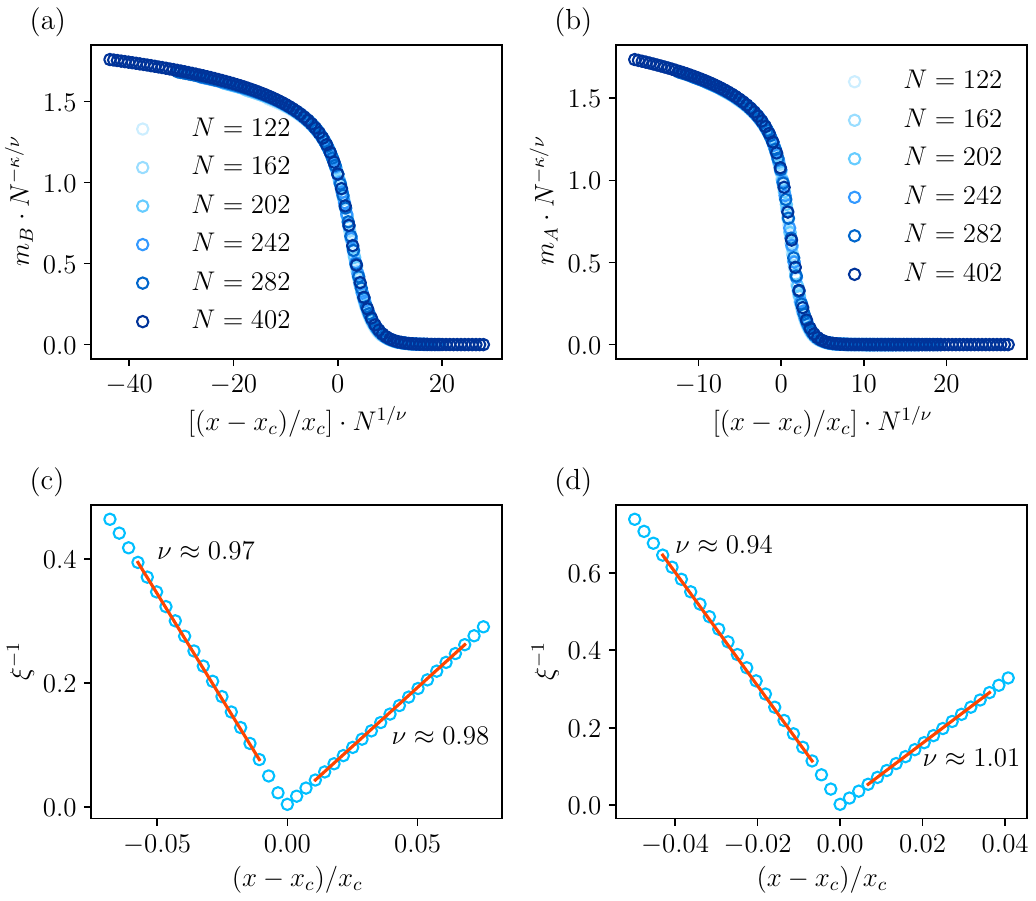}
    \caption{
    (a) Data collapse of order parameter $m_B$ by using $x_c=2.805$, $\nu=1$ and $\kappa=1/8$. 
    (b) Data collapse of order parameter $m_A$ by using $x_c=4.446$, $\nu=1$ and $\kappa=1/8$. 
    (c) and (d) The $\nu$ are extracted by using $\xi\sim|x-x_c|^{-\nu}$ and the $\xi$ are calculated from VUMPS. 
    The $x_c$ is chosen to be the point with maximum correlation length, for (c) $x_c=2.79$ and $x_c=4.42$ for (d). 
    \label{fsa1}
    }
\end{figure}
Here we first provide finite-size scaling analysis of the consecutive quantum phase transitions 
along $R_b/d=1.3$ in Fig.~\ref{phase-diagram}(a). 
Our conclusion is that both quantum phase transitons seperating order-disorder mixture phase 
from ordered and disordered phases belong to the Ising universality class. 
We define two order parameters for A atoms and B atoms respectively, 
\begin{eqnarray}
    m_A&=&\frac{2}{N}\sum_{i=1}^{N/2}(-1)^{i+1}n_{2i-1} \\
    m_B&=&\frac{2}{N}\sum_{i=1}^{N/2}(-1)^{i+1}n_{2i}
\end{eqnarray}
and according to the finite-size scaling laws for a singular quantity $Q$ we have 
\begin{equation}
    Q(\tilde{x}, N)=N^{\kappa/\nu}g(\tilde{x}N^{1/\nu})
\end{equation}
where $\tilde{x}=(x-x_c)/x_c$ is the normalized model parameter, 
$N$ is the system size and $g(\cdot)$ is the scaling function. 
If the scaling exponent is correct, data for different (large) system sizes should fall 
onto the same curve, which then is the scaling function \cite{FiniteSizeScaling}. 

The curves of the order parameters obtained from DMRG calculations 
for systems of different sizes exhibit a clear data collapse when 
rescaled using the critical exponents $\kappa = -\beta = -1/8$ and $\nu = 1$, 
as shown in Fig.~\ref{fsa1}(a) and (b), corresponding to the two phase transitions 
near $x_c \approx 2.8$ and $x_c \approx 4.4$ in Fig.~\ref{phase-diagram}(b). 
To further substantiate that both transitions belong to the Ising universality class, 
we perform additional calculations using the VUMPS algorithm to extract the 
correlation length and numerically determine the critical exponent $\nu$. 
The fitted values of $\nu$, presented in Fig.~\ref{fsa1}(c) and (d), are 
found to be close to the theoretical value $\nu=1$ for both transitions, 
providing strong evidence that they are indeed of the Ising type.
\begin{figure}
    \includegraphics[scale=0.52]{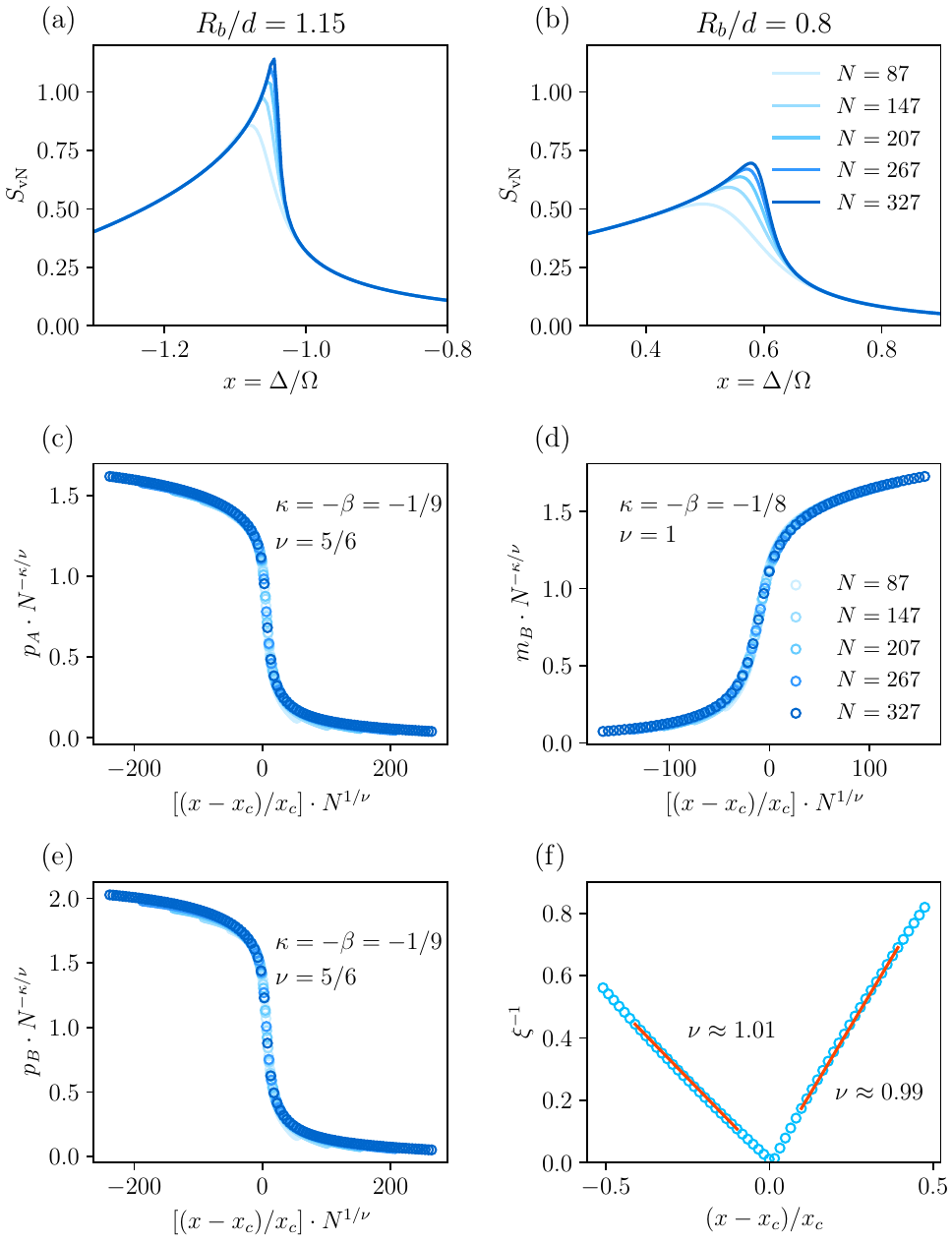}
    \caption{
    (a)(b) Bipartite entanglement entropy $S_{\mathrm{vN}}$ as the function of $x=\Delta/\Omega$ are shown over different 
    system size $N$. 
    (c)(d)(e) Data collapse of order parameters $p_A,p_B$ and $m_B$. 
    (f) For the phase transition between disordered and half-ordered phases, the correlation length are 
    calculated to extract critical exponent $\nu$. 
    \label{fsa2}
    }
\end{figure}
We also provide the data collapse of the two second-order phase 
transitions shown in Fig.~\ref{phase-diagram-2}. 
The order parameter $p_A, p_B$ are defined to be 
\begin{eqnarray}
    p_A&=&\left|\frac{3}{N}\sum_{i=1}^{\lceil N/2\rceil} e^{i2\pi(i-1)/3}n_{2i-1} \right|\\
    p_B&=&\left|\frac{3}{N}\sum_{i=1}^{\lfloor N/2\rfloor} e^{i2\pi(i-1)/3}n_{2i}\right|.
\end{eqnarray}
The data collapse of $p_A$ and $p_B$ is shown in Fig.~\ref{fsa2}(c) and (d) with 
critical exponents $\nu=5/6$ and $\beta=1/9$. Thus the phase transiton between disordered and ordered phases 
belongs to the 3-state Potts universality class. 
Similarly, the data collapse of order paramter $m_B$ shown in Fig.~\ref{fsa2}(d), together 
with the extracted correlation length shown in Fig.~\ref{fsa2}(f), proves the phase transition 
between disordered and half-ordered phases is also of Ising type.

\section{Details on the calculation of level spacings}
\begin{figure}
    \includegraphics[scale=0.223]{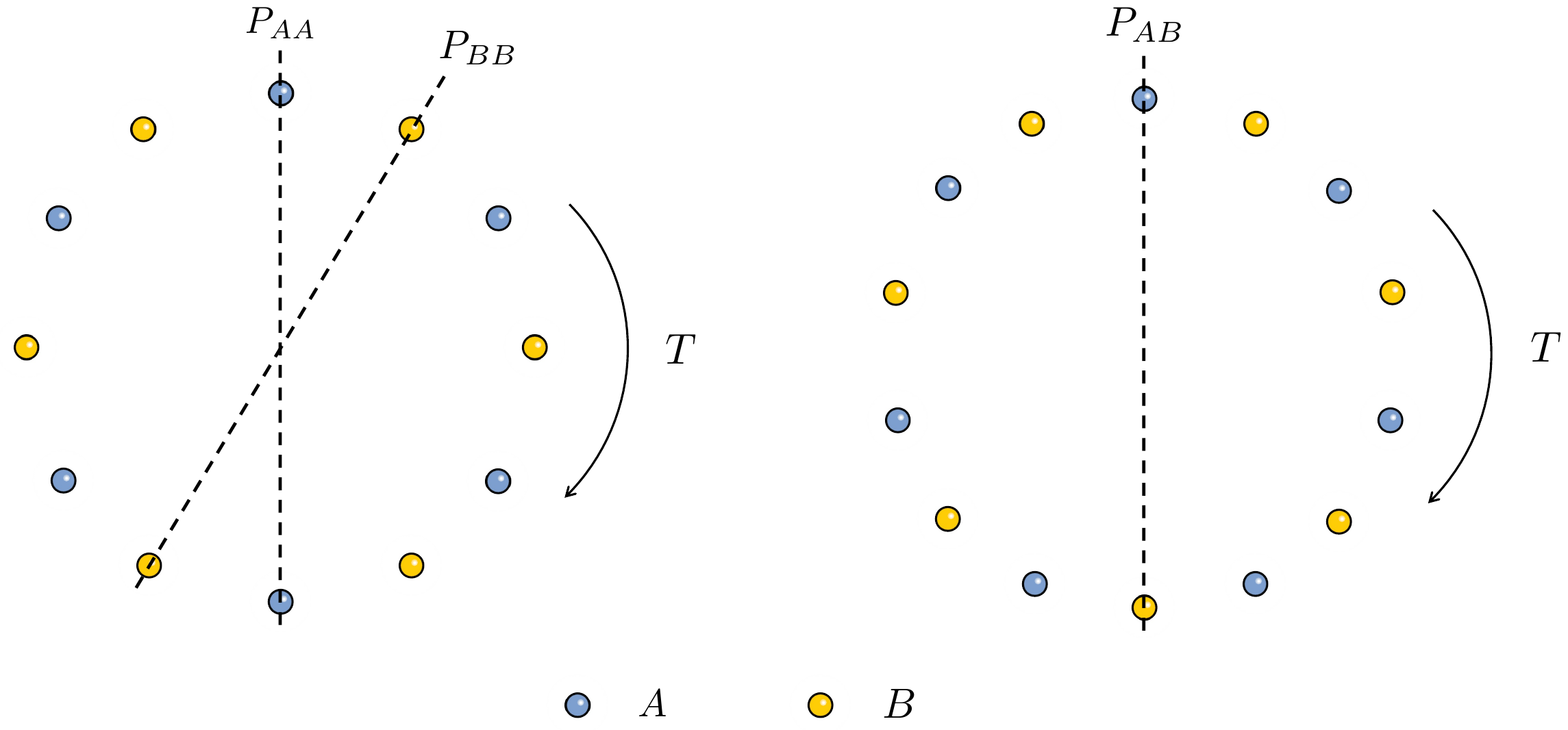}
    \caption{
    Symmetry operations in the two-species Rydberg atoms array under PBC 
    for system size $N = 4n, n\in\mathbb{N}^{+}$(left) and 
    $N = 4n+2, n\in\mathbb{N}^{+}$(right) are depicted above where the 
    blue(yellow) circle stands for type A(B) atom. 
    \label{symmetry-operation}
    }
\end{figure}

There are three(two) kinds of symmetry operations in the atom array with 
system $N = 4n, n\in\mathbb{N}^{+}$($N=4n+2, n\in\mathbb{N}^{+}$) 
denoted $\hat{T}$, $\hat{P}_{AA}$ and $\hat{P}_{BB}$($\hat{T}, \hat{P}_{AB}$) 
respectively. Their physical meaning is depicted in Fig.~\ref{symmetry-operation}.

We focus on the $N = 4n$ ($n\in\mathbb{N}^+$) system. To calculate the level spacing distribution, we have 
to reduce all the symmetries in the system. For the caculation in the main text, we choose to 
diagonalize the system Hamiltonian in $k = 0$ and $p_{AA}=p_{BB}=+1$ sector, i.e., translational 
invariant and inversion even sector. One can note that these three symmetry operations 
do not commute with one another indicating they do not have common eigenstates. However, 
it can be proved that within the $k = 0$ subspace the three symmetry operations all commute or, 
more formally, all the translational invariant states 
\begin{equation}
    \ket{k_a} = \frac{1}{\sqrt{N_a}}\sum_{r=0}^{K-1}\hat{T}^{r}\ket{a}
\end{equation}
fall into the kernal of the commutator of arbitary two operations out of three, 
where we use $\ket{k_a}$ to denote the momentum zero eigenstates, $\ket{a}$ to 
denote product states in computational basis, $K = N/2$ to denote the number of unit cell and 
$\sqrt{N_a}$ is the normalization coefficient.

The proof is quite direct. Considering an arbitrary computational basis $\ket{a} = |i_{0}i_{1}i_{2}\cdots i_{N-1}\rangle$, 
the outcome states acted on by symmetry operations $\hat{T}, \hat{P}_{AA}, \hat{P}_{BB}$ are 
\begin{eqnarray}
    \hat{T}\ket{i_{0}i_{1}i_{2}\cdots i_{N-1}} &=& \ket{i_{N-2}i_{N-1}i_0\cdots i_{N-3}}\\
    \hat{P}_{AA}\ket{i_{0}i_{1}i_{2}\cdots i_{N-1}} &=& \ket{i_0 i_{N-1}i_{N-2}\cdots i_1}\\
    \hat{P}_{BB}\ket{i_{0}i_{1}i_{2}\cdots i_{N-1}} &=& \ket{i_2 i_1 i_0 i_{N-1}\cdots i_{3}}
\end{eqnarray}
respectively from which it is easy to obtain the following relation
\begin{eqnarray}
    \hat{P}_{AA}\hat{T} &=& \hat{T}^{-1}\hat{P}_{AA}\\
    \hat{P}_{BB}\hat{T} &=& \hat{T}^{-1}\hat{P}_{BB}\\
    \hat{T} &=& \hat{P}_{BB}\hat{P}_{AA}.
\end{eqnarray}
Therefore, for any translational invariant states, one can find
\begin{eqnarray}
    [\hat{P}_{AA}, \hat{T}]\ket{k_a} &=& \hat{P}_{AA}\hat{T}\ket{k_a} - \hat{T}\hat{P}_{AA}\ket{k_a} \\
     &=& \hat{P}_{AA}\hat{T}\ket{k_a} - \hat{P}_{AA}\hat{T}^{-1}\ket{k_a} = 0 
\end{eqnarray}
and similarly $[\hat{P}_{BB}, \hat{T}]\ket{k_a}=0$ where we use $\hat{T}\ket{k_a} = \hat{T}^{-1}\ket{k_a} = \ket{k_a}$ because 
$\{\ket{k_a}\}$ are translational invariant states. Due to the same reason, the following equation is also true.
\begin{eqnarray}
    [\hat{P}_{AA}, \hat{P}_{BB}]\ket{k_a} = \hat{T}^{-1}\ket{k_a}-\hat{T}\ket{k_a}=0
\end{eqnarray}
So by symmetrizing as follows
\begin{equation}
    \ket{k_a,p_{AA},p_{BB}} = \frac{1}{\mathcal{N}_a}(\mathbb{I}+\hat{P}_{AA})(\mathbb{I}+\hat{P}_{BB})\ket{k_a}
\end{equation}
we can get the basis states in $k=0$ and $p_{AA}=p_{BB}=+1$ sector where 
$1/\mathcal{N}_a$ is only a normalization factor. Similarly, it can be 
concluded that $[\hat{P}_{AB}, \hat{T}]\ket{k_a}=0$ in the $N=4n+2$ system.

For system with $N = 16$, there are $4371$ basis states in the $k=0, p_{AA}=p_{BB}=+1$ symmetry sector after which 
the symmetry reduced Hamiltonian $H_{r} = UHU^{\dagger}$ can be directly 
decompose to obtain the eigenvalues where each column of $U$ corresponds to a specific $\ket{k_a}$ 
in the computational basis.

\section{Constructions of approximate quantum many-body scars}

\begin{figure}
    \includegraphics[scale=0.378]{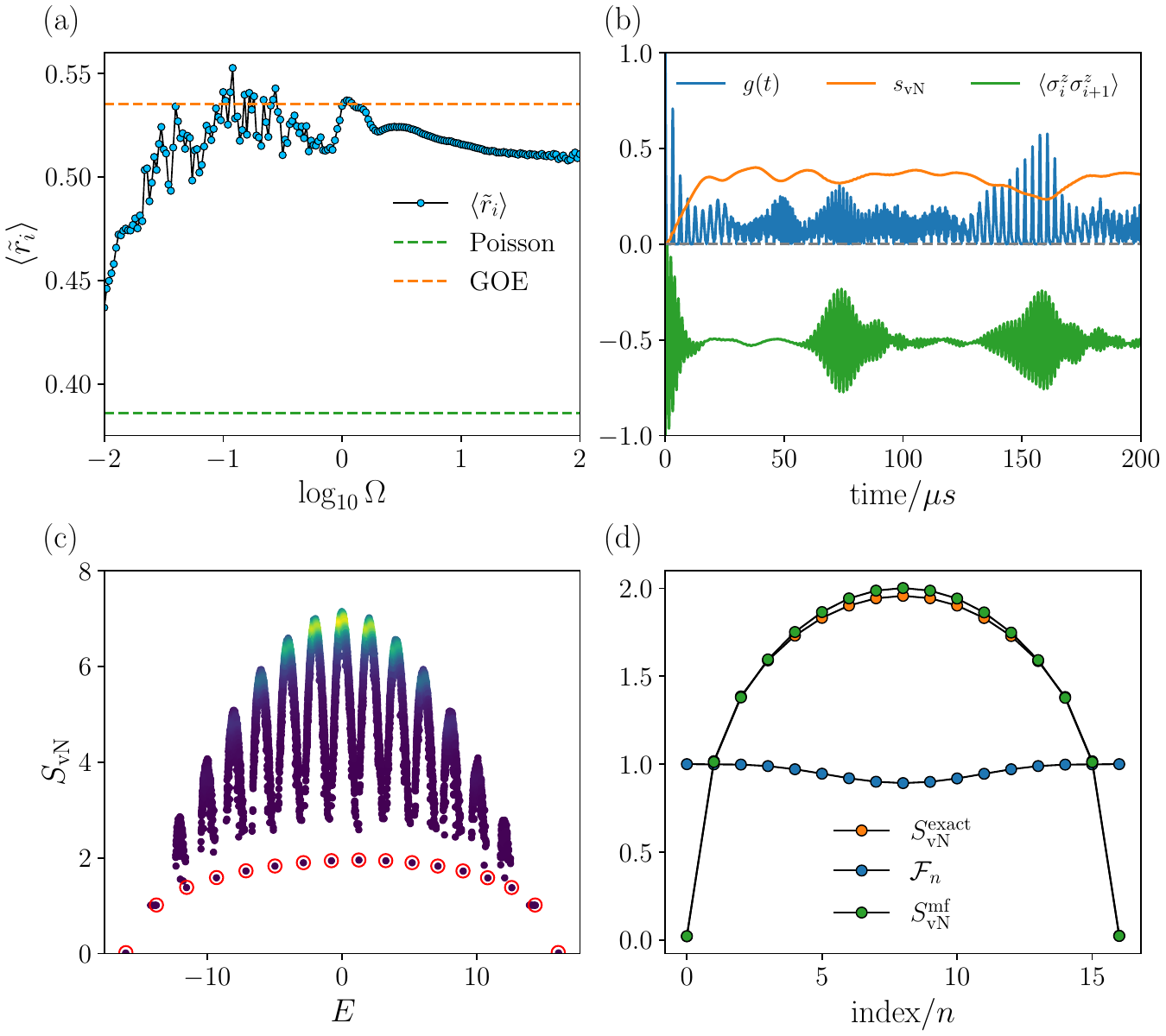}
    \caption{
    Numerical results for ANNNI model with system size $N=16$ 
    and coefficients in Hamiltonian \eqref{ANNNI-H} 
    are $(\Omega, V_1, V_2)=(2, -0.062, 0.074)$. 
    (a) The level spacing indicator $\langle \tilde{r}_i\rangle$ 
    as a function of $\log_{10} \Omega$ in the symmetry 
    reduced sector reveals that when $\Omega \gtrsim 0.1$ the system 
    behaves in a chaotic way even if $\Omega$ is very large ($\Omega>10$). 
    (b) Quench dynamics of the ANNNI model 
    from $\ket{\mathbb{Z}_2}$ also shows non-thermaling behavior 
    similar to the two-species Rydberg atoms array. 
    (c) The eigenstates' energy-entanglement spectrum is similar to the two-species Rydberg atoms array, 
    where they both unfold low bipatite entanglement entropy eigenstates with the same arch structure. 
    (d) Fidelity and bipartite entanglement entropy of the states constructed 
    by the quasiparticle operator $\hat{\beta}_k^{\dagger}$ are compared with the exact 
    eigenstates obtained via numerical method. Here, the fidelity $\mathcal{F}_n$ is defined as 
    the overlap between exact state $\ket{s_n}$ and constructed state $\ket{s_n'}$, i.e., 
    $\mathcal{F}_n = |\langle s_n|s_n'\rangle|^2$. 
    The bipartite entanglement entropy of the mean 
    field states $S_{\mathrm{vN}}^{\mathrm{mf}}$ are 
    very close to the entropy of the exact results $S_{\mathrm{vN}}^{\mathrm{exact}}$.
    \label{ANNNIfig}
    }
\end{figure}

In the theoretical consideration, we only retain interactions between atoms separated by up to two atom 
spacings, which means only the nearest and next nearest interactions will be taken into consideration. 
We first map the Rydberg system Hamiltonian \eqref{Hamiltonian} to the quantum Ising model Hamiltonian using 
$\hat{n}_i = (\mathbb{I}+\hat{\sigma}_i^z)/2$ and then one obtains Eq.~\eqref{C1}, 
\begin{widetext}
\begin{eqnarray}
    \hat{H}= \frac{\Omega}{2}\left(\sum_{i}\hat{\sigma}_{2i}^x+\hat{\sigma}_{2i+1}^x\right) + 
    \frac{V_{AA}}{4}\sum_{i}\hat{\sigma}_{2i}^z\hat{\sigma}^z_{2i+2}
    +\frac{V_{BB}}{4}\sum_{i}\hat{\sigma}_{2i+1}^z\hat{\sigma}^z_{2i+3} 
    +\frac{V_{AB}}{4}\left(\sum_{i}\hat{\sigma}_{2i}^z\hat{\sigma}^z_{2i+1}+\hat{\sigma}_{2i+1}^z\hat{\sigma}^z_{2i+2}\right)\nonumber\\
    -\frac12(\Delta-V_{AA}-V_{AB})\sum_{i}\hat{\sigma}_{2i}^z-\frac12(\Delta-V_{BB}-V_{AB})\sum_{i}\hat{\sigma}_{2i+1}^z 
    + \frac{N}{8}(V_{AA}+V_{BB}+2V_{AB}-4\Delta)\mathbb{I}.\label{C1}
\end{eqnarray}
\end{widetext}
where $V_{AA}=C_{AA}/(2d)^6$, $V_{BB}=C_{BB}/(2d)^6$ and $V_{AB} = C_{AB}/d^6$. To further 
simplify the Hamiltonian, we take some approximations. From now on, we only   
consider the situation where $V_{AA}=V_{BB}$ and the $\Delta$ is approriately chosen such that single $\hat{\sigma}^z_{i}$ term 
vanishes, i.e., $\Delta - V_{AA}-V_{AB}=0$ and $\Delta - V_{BB}-V_{AB}=0$. Therefore, the two-species Rydberg atoms array Hamiltonian 
is simplified into the quantum ANNNI model Hamiltonian. We denote the nearest interaction strength $V_1=V_{AB}/4$ as well as the  
next nearest interaction strength $V_2 = V_{AA}/4=V_{BB}/4$, and then the simplified Hamiltonian writes
\begin{equation}
    \hat{H}_{\mathrm{ANNNI}} = \frac{\Omega}{2}\sum_n\hat{\sigma}^x_n+V_1\sum_n\hat{\sigma}^z_{n}\hat{\sigma}^z_{n+1}+
    V_2\sum_n\hat{\sigma}_n^z\hat{\sigma}^z_{n+2} \label{ANNNI-H}
\end{equation}
where the integrability breaks due to the next nearest interaction term which can be 
proved by the level spacing indicator $\langle\tilde{r}_i\rangle$ in Fig.~\ref{ANNNIfig}(a). 

First, we perform local unitary transformation followed 
by the Jordan-Wigner transformation such that 
$\hat{\sigma}^x\leftrightarrow\hat{\sigma}^z$ and the same Hamiltonian written in 
spinless fermion operators can be obtained (Eq.~\eqref{JWH}). 
\begin{eqnarray}
    & &\hat{H}_{\mathrm{JW}} = \Omega\sum_n
    \hat{c}_n^{\dagger}\hat{c}_n+V_1\sum_n(
    \hat{c}_n-\hat{c}_n^{\dagger})(\hat{c}_{n+1}^{\dagger}+\hat{c}_{n+1}) \nonumber \\
    & &+V_2\sum_n(\hat{c}_n-\hat{c}_n^{\dagger})(2\hat{c}_{n+1}^{\dagger}\hat{c}_{n+1}-1)
    (\hat{c}_{n+2}^{\dagger}+\hat{c}_{n+2})
    \label{JWH}
\end{eqnarray}
Second, mean-field approximation can be applied. We define
\begin{eqnarray}
    \alpha &=& \langle 2\hat{c}_{n+1}^{\dagger}\hat{c}_{n+1}-1\rangle=\braket{\hat{\sigma}^z_{n+1}} \label{defa}\\
    \mu &=& \langle (\hat{c}_n^{\dagger}-\hat{c}_n)(\hat{c}_{n+2}^{\dagger}+\hat{c}_{n+2})\rangle \label{defmu}\\
    &=&\braket{\hat{\sigma}^x_{n}\hat{\sigma}^z_{n+1}\hat{\sigma}^x_{n+2}}
\end{eqnarray}
and denote $\tilde{\Omega}=\Omega+2\mu V_2$ as well as 
$\tilde{V}_2=\alpha V_2$, and then the mean-field Hamiltonian $\hat{H}_{\mathrm{MF}}$ reads (ignoring a constant)
\begin{eqnarray}
    \hat{H}_{\mathrm{MF}} &=& \tilde{\Omega}\sum_n
    \hat{c}_n^{\dagger}\hat{c}_n+V_1\sum_n(\hat{c}_n-\hat{c}_n^{\dagger})(\hat{c}_{n+1}^{\dagger}+\hat{c}_{n+1}) \nonumber \\
    & &+\tilde{V}_2\sum_n(\hat{c}_n-\hat{c}_n^{\dagger})(\hat{c}_{n+2}^{\dagger}+\hat{c}_{n+2})
\end{eqnarray}
where it has to be emphasized that the mean-field Hamiltonian $\hat{H}_{\mathrm{MF}}$ possesses 
fermion parity conservation which naturally gives rise to 
the two different symmetry sectors with even and odd parity. 
Different symmetry sectors correspond with different boundary conditions. 
The even(odd) parity $p=0$ ($p=1$) is associated with the PBC(ABC), which is clearly shown in Eq.\eqref{bc}, 
\begin{eqnarray}
    \hat{c}_{N} = (-1)^{p} \hat{c}_0 \label{bc}
\end{eqnarray}
where we use 0 to label the first atom. 
The $p=0$ ($p=1$) sector indicates $e^{ikN}=1$ ($e^{ikN}=-1$) and thus the choice of wave vector $k$ given by 
PBC (ABC) is $\mathcal{K}^{\mathrm{ABC}}_{p=0}$ ($\mathcal{K}^{\mathrm{PBC}}_{p=1}$)
\begin{eqnarray}
    \mathcal{K}^{\mathrm{PBC}}_{p=0}&=&\left\{k=\frac{2m\pi}{N}, m = -\frac{N}{2}+1,\cdots, \frac{N}{2}\right\}\\
    \mathcal{K}^{\mathrm{ABC}}_{p=1}&=&\left\{k=\pm\frac{(2m-1)\pi}{N}, m = 1,\cdots, \frac{N}{2}\right\} 
\end{eqnarray}
where $m\in\mathbb{Z}$.
Next, we obtain the energy spectrum using Fourier transformation.
\begin{eqnarray}
    \hat{c}_n&=&\frac{1}{\sqrt{N}}\sum_k e^{-ikn}\hat{c}_k\\
    \hat{c}_n^{\dagger}&=&\frac{1}{\sqrt{N}}\sum_k e^{ikn}\hat{c}_k^{\dagger}
\end{eqnarray}
After defining
\begin{eqnarray}
    A &=& \tilde{\Omega}-2V_1\cos k-2\tilde{V}_2\cos 2k\\
    B &=& 2(V_1\sin k+\tilde{V}_2\sin 2k)\\
    C &=& \sqrt{A^2+B^2}
\end{eqnarray}
the Hamiltonian in momentum space can be written in a simple form
\begin{equation}
    \hat{H}_{\mathrm{MF}} = \sum_k\begin{pmatrix}
    \hat{c}_k^{\dagger} & \hat{c}_{-k}\end{pmatrix}
    \begin{pmatrix}
        A & iB\\
        -iB & -A
    \end{pmatrix}\begin{pmatrix}
        \hat{c}_k \\ \hat{c}^{\dagger}_{-k}
    \end{pmatrix}
\end{equation}
and the dipersion relation is $\epsilon_k = \pm C$.

To construct the quasipaticle operator $\hat{\beta}_k^{\dagger}$, we use the Bogoliubov 
transformation 
\begin{equation}
    \begin{pmatrix}
        \hat{\beta}_k \\ \hat{\beta}_{-k}^{\dagger}
    \end{pmatrix} = \begin{pmatrix}
        u_k & -v_k\\
        v_k^* & u_k^*
    \end{pmatrix}\begin{pmatrix}
        \hat{c}_k \\ \hat{c}_{-k}^{\dagger}
    \end{pmatrix}
\end{equation}
where $u_k$ is real and $v_k$ is purely imaginary
\begin{equation}
    u_k=\frac{A+C}{\sqrt{2C(A+C)}};\ v_k=\frac{-iB}{\sqrt{2C(A+C)}}
\end{equation}
and the Hamiltonian in the language of quasiparticle operators reads
\begin{equation}
    \hat{H}_{\mathrm{MF}} = \sum_k\epsilon_k(\hat{\beta}_k^{\dagger}\hat{\beta}_k+\hat{\beta}_{-k}^{\dagger}\hat{\beta}_{-k}-1)
\end{equation}
from which the ground state can be directly deduced, i.e., the Bogoliubov vacuum $\ket{\emptyset}$ satisfying
$\hat{\beta}_k\ket{\emptyset} = 0, \forall k$. Here, we need to figure out which symmetry sector 
the real ground state falls into because there are two ground states, one ($\ket{\emptyset}^{\mathrm{ABC}}$) in the 
ABC sector and the other ($\ket{\emptyset}^{\mathrm{PBC}}$) in the PBC sector. 
It turns out that the real ground state falls into the ABC sector, indicating $\ket{\emptyset}^{\mathrm{ABC}}$ is 
the ground state of the whole system. The two ground states in terms of fermoin operators can be expressed
\begin{eqnarray}
    \ket{\emptyset}^{\mathrm{ABC}} &=& 
    \prod_{k>0}^{\mathcal{K}^{\mathrm{ABC}}_{p=1}}\Big(u_k+v_k\hat{c}_k^{\dagger}\hat{c}_{-k}^{\dagger}\Big)
    \ket{0}\\
    \ket{\emptyset}^{\mathrm{PBC}} &=& \hat{c}_{k=0}^{\dagger}
    \prod_{0<k<\pi}^{\mathcal{K}^{\mathrm{PBC}}_{p=0}}\Big(u_k+v_k\hat{c}_k^{\dagger}\hat{c}_{-k}^{\dagger}\Big)
    \ket{0}
\end{eqnarray}
where $\ket{0}$ is the fermion vacuum $\hat{c}_k\ket{0} = 0, \forall k$ 
and $\hat{c}_{k=0}^{\dagger}$ is the creation operator 
of a fermion with zero momentum.

We found that the low bipartite entanglement entropy states $\ket{s_n}$ can be 
well contructed using states in Eq. \eqref{quasiparicle-state} with $\sum_k n_k = \lfloor n/2\rfloor$.
Specifically, we use $\ket{\psi_{\{n_k\}}^{\mathrm{ABC}/\mathrm{PBC}}}$ as the basis and numerically calculate 
the coefficient $c_{\{n_k\}}=\langle\psi_{\{n_k\}}^{p=n\ \mathrm{mod}\ 2}|s_n\rangle$ in Eq. \eqref{C21}, 
\begin{equation}
    \ket{s'_n} = \frac{1}{\mathcal{N}}\sum_{\{n_k\}}^{\sum_k n_k = \lfloor n/2\rfloor} 
    c_{\{n_k\}}\ket{\psi_{\{n_k\}}^{p=n\ \mathrm{mod}\ 2}}\label{C21}
\end{equation}
where $\mathcal{N}$ is the normalization coefficent, $\ket{\psi_{\{n_k\}}^{p=0}}=\ket{\psi_{\{n_k\}}^{\mathrm{ABC}}}$ and 
$\ket{\psi_{\{n_k\}}^{p=1}}=\ket{\psi_{\{n_k\}}^{\mathrm{PBC}}}$. 

In the following, we show some the numerical results. 
The ANNNI model has the similar eigenstate energy-entanglement spectrum with 
the two-species Rydberg atoms array which can be clearly seen in Fig.~\ref{ANNNIfig}(c), 
where those dots marked with red circles are regarded as the approximate quantum many-body scars, 
which can be well captured by the basis shown in Eq.~\eqref{C21}. 
For system of small size ($N=16$), the calculation results shows that the fidelity 
$\mathcal{F}_n = |\langle s_n'|s_n\rangle|^2 >0.89$ when the coefficients in the 
Hamiltonian are chosen to be $(\Omega, V_1, V_2)=(2, -0.062, 0.074)$ and 
the variational parameters are chosen to be $(\alpha, \mu) = (2n/N-1, 0)$. 
The fidelity $\mathcal{F}_n$ can also be transformed to the 
the average value of zero-momentum subspace projector $\hat{P}_{k=0}$ and we find 
\begin{equation}
    \langle s_n|\hat{P}_{k=0}|s_n\rangle > 0.796
\end{equation}
where 
\begin{eqnarray}
    \hat{P}_{k=0} &=& \sum_{\{n_k\}}^{\sum_k n_k = \lfloor n/2\rfloor} \hat{P}_{\{n_k\}}\\
    \hat{P}_{\{n_k\}} &=& \ket{\psi_{\{n_k\}}^{p=n\ \mathrm{mod}\ 2}}
    \langle \psi_{\{n_k\}}^{p=n\ \mathrm{mod}\ 2}|.
\end{eqnarray}

We note that one can optimize these mean-field 
variational parameters $(\alpha, \mu)$ to obtain higher 
fidelity.

\section{Discussion on the possible DQPT phenomenon}

\begin{figure}
    \includegraphics[scale=0.4]{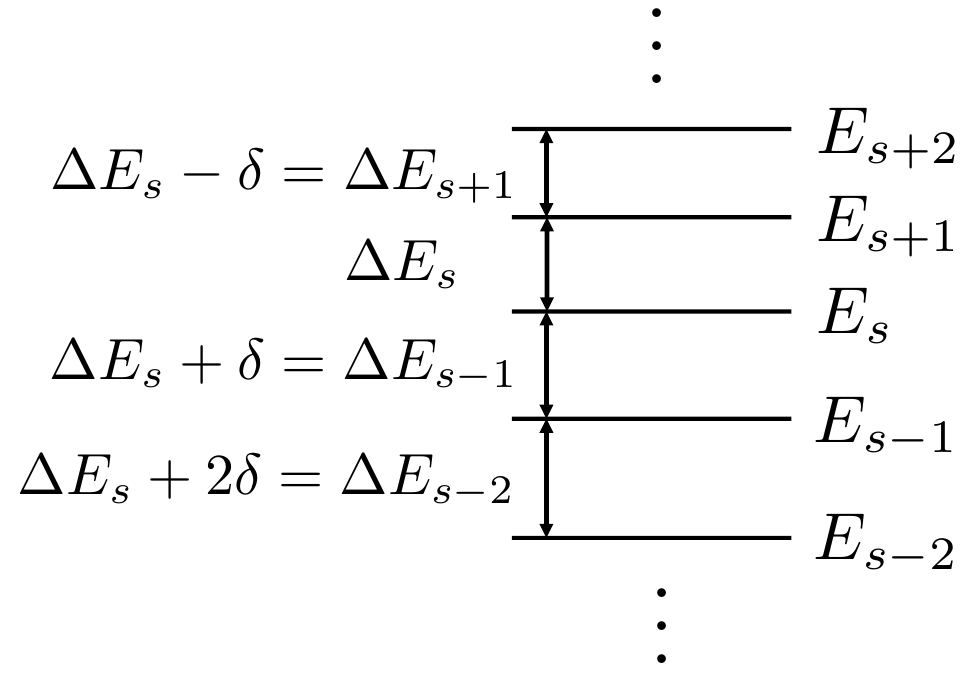}
    \caption{
    The energy levels of scar states exhibit decreasing energy 
    level spacing with growing index $s$ or energy $E_s$. Here we define 
    $\Delta E_s = E_{s+1}-E_s$.
    \label{energy_level}
    }
\end{figure}

\begin{figure}
    \includegraphics[scale=0.53]{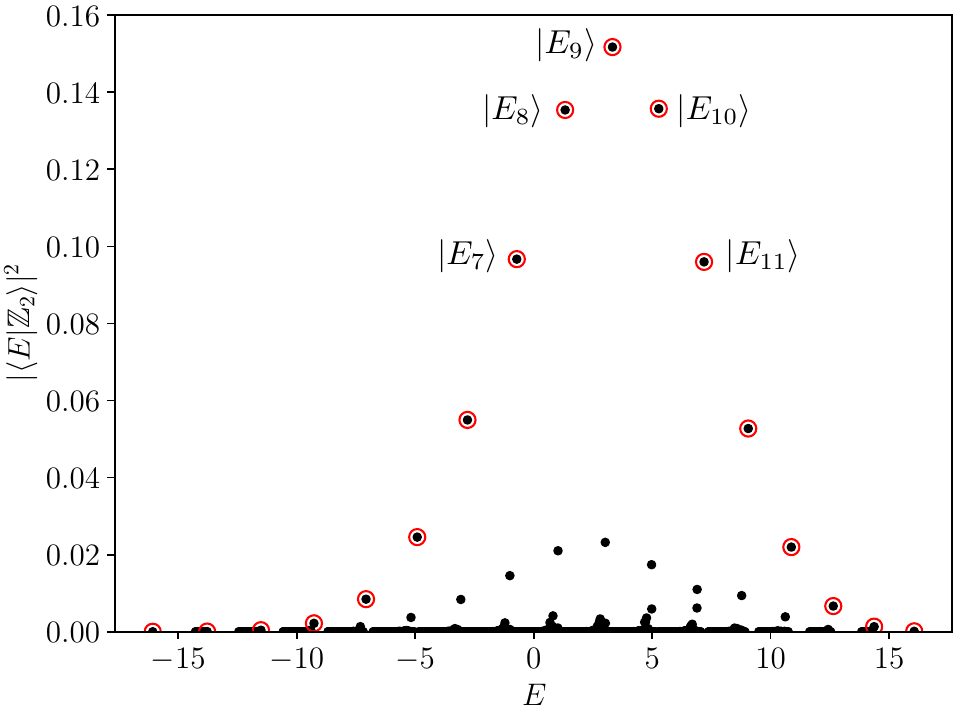}
    \caption{
    The overlap of $\ket{\mathbb{Z}_2}$ with all eigenstates is depicted where those marked with 
    red circles are the ones with approximate scar states. 
    In the main text and Fig.~\ref{quantum-quench}, the $\lambda_S(t)$ is derived from 
    $\ket{E_i}, i=7\sim11$ while $\tilde{\lambda}_S(t)$ incorporates all scarred eigenstates marked by red circles.
    \label{scar_overlap}
    }
\end{figure}

\begin{figure}
    \includegraphics[scale=0.56]{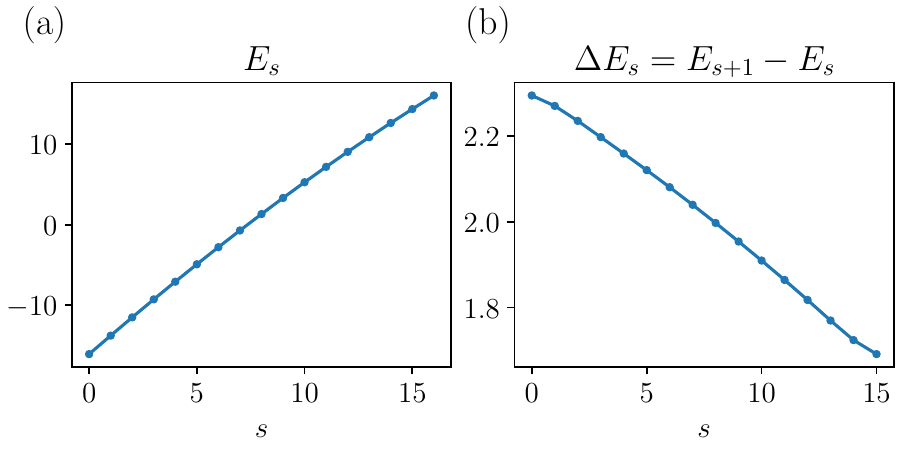}
    \caption{
    The eigenenergy of scar states $E_s$ and their neighbouring spacings 
    $\Delta E_s = E_{s+1}-E_s$ are depicted in (a) and (b) respectively. 
    The decreasing energy level spacings are visulized in Fig.~\ref{energy_level}.
    \label{energy_level_spacing}
    }
\end{figure}

\begin{figure*}
    \includegraphics[scale=0.585]{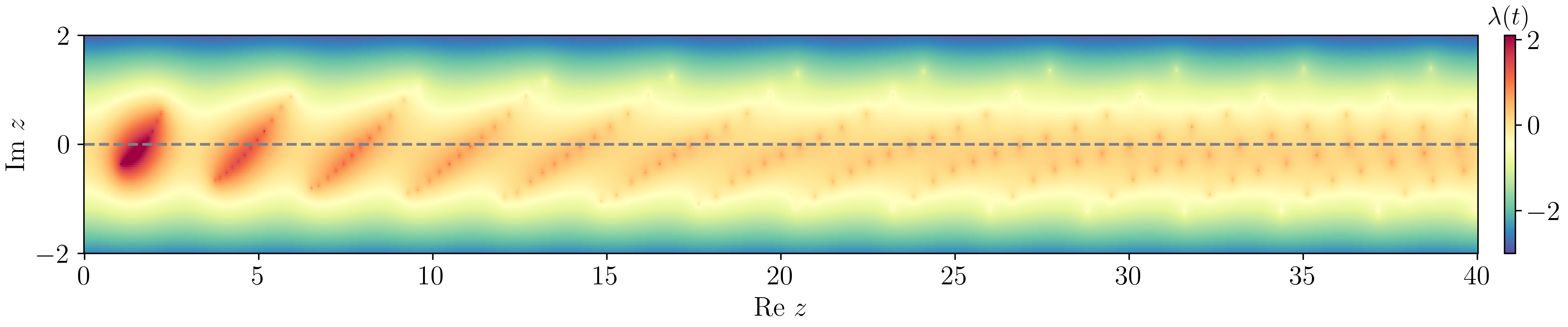}
    \caption{
    Loschmidt rate $\lambda(z)=-\frac{1}{N}|\langle \mathbb{Z}_2|e^{-i\hat{H}z}|\mathbb{Z}_2\rangle|^2$ over the complex plane is calculated for Rydberg atoms array with size $N=12$ where 
    $(\Omega, \Delta, d) = (2\ \mathrm{MHz}, 0.05\ \mathrm{MHz}, 9\ \mu m)$ and the $C_6$ coefficients are chosen from 
    the first line of Table \ref{interaction-data}. Those red dots around real axis are 
    the possible zeros of Loschmidt amplitude in the thermodynamic limit and their number 
    corresponds to the number of scar states in the system. 
    Grey dashed line is the real axis and also the evolution path of the system.
    \label{DQPT_rate_complex_plane}
    }
\end{figure*}

\begin{figure}
    \includegraphics[scale=0.425]{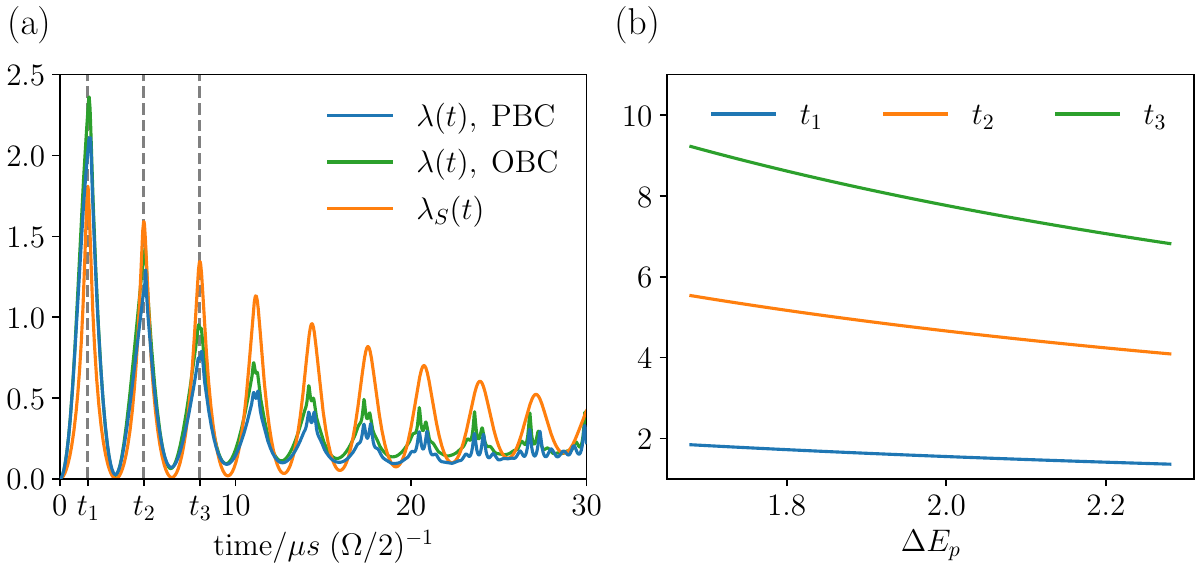}
    \caption{
        Discussion on the non-analytical behavior of Loschmidt rate $\lambda(t)$. 
        (a) The Loschmidt rate of $N = 16$ system is calculated for both PBC and OBC and 
        $\lambda_S(t)$ constructed using our bold approximations is also depicted here. It 
        can be found that our $\lambda_S(t)$ corresponds well with the exact 
        $\lambda(t)$ obtained via ED. 
        (b) The first three time points $t_1, t_2, t_3$ where approximate Loschmidt rate $\lambda_S(t)$ shows non-analytical 
        behavior are marked with grey dashed line in (a) and their relation with $\Delta E_p$ is 
        depicted here.
        \label{dqpt_rate_s}
    }
\end{figure}

We use numerical integration scheme to calculate the 
quench dynamics for small system and TDVP method for larger system. 
In both cases, the discontinous behavior of the Loschmidt rate can be 
easily identified as shown in Fig.~\ref{quantum-quench}(f).

Generally speaking, DQPT can be explained by the Lee-Yang zeros (also called Fisher zeros sometimes) of 
the Loschmidt amplitude $\mathcal{G}(z)=\bra{\psi_0}e^{-i\hat{H}z}\ket{\psi_0}$ in the complex plane, 
where $\ket{\psi_0}$ is the initial state , $z=t+i\tau$ is the complex time and we choose 
$\ket{\psi_0}=\ket{\mathbb{Z}_2}$ here. When 
$\mathcal{G}(z)\to 0$, the Loschmidt rate $\lambda(z)=-\frac{1}{N}\ln|\mathcal{G}(t)|^2$ will 
diverge and that is why $\lambda(t)$ becomes discontinous at some 
point. We calculated the $\lambda(z)$ over the complex plane as 
shown in Fig.~\ref{DQPT_rate_complex_plane} and found that those 
possible Lee-Yang zeros tend to form lines crossing the real axis with the system becoming larger 
and their number corresponds with the number of QMBS in the system. 
This phenomenon can be partially explained by the approximate QMBS we discussed above. 

The initial state $\ket{\mathbb{Z}_2}$ can be formally decomposed by the eigenstates $\ket{E_k}$ and also can be 
approximately decomposed by the scar states $\ket{E_s}$ where we use $k$ to label general eigenstates and $s$ to label scar eigenstates, 
\begin{equation}
    \ket{\mathbb{Z}_2} = \sum_{k=0}^{2^N-1} c_k\ket{E_k}\approx \sum_{s=0}^{N} c_s\ket{E_s}
\end{equation}
so the Loschmidt amplitude is 
\begin{equation}
    \mathcal{G}(t) = \bra{\mathbb{Z}_2}e^{-i\hat{H}t}\ket{\mathbb{Z}_2}
    =\sum_s|c_s|^2 e^{-iE_st}
\end{equation}
and the corresponding Loschmidt echo $\mathcal{L}(t) = |\mathcal{G}(t)|^2$ is 
\begin{equation}
    \mathcal{L}(t) = \sum_{s,s'}|c_s|^2|c_{s'}|^2e^{-i(E_s-E_{s'})t}. \label{E3}
\end{equation}
The scar states in our system are approximiate so their energy level spacings are not equal, 
we consider the situation where $(E_{s+1}-E_s)-(E_s-E_{s-1}) = \delta$ is a constant 
as shown in Fig.~\eqref{energy_level}.

To further simplify our discussion, we now only consider the contribution of 
five scar states $\{\ket{E_s}, s=7, 8, 9, 10, 11\}$ (taking $N=16$ as an example here) 
to the system dynamics because they have 
the largest overlap with $\ket{\mathbb{Z}_2}$ which can be confirmed in Fig.~\ref{scar_overlap}.
Hence the Eq.~\eqref{E3} approximately equals to 
\begin{equation}
    \mathcal{L}(t)\approx\sum_{s'=7}^{10}\sum_{s=s'}^{11}2|c_s|^2|c_{s'}|^2\cos[(E_s-E_{s'})t] \label{E4}
\end{equation}
and from now on we will make bold approximations. We only consider the low frequency part of the 
Loschmidt echo in Eq.~\eqref{E4}, which leads to the final expression for Loschmidt echo $\mathcal{L}_S(t)$
\begin{equation}
    \mathcal{L}_S(t)=\sum_{s=7}^{10}2\Big( |c_{s}|^2|c_{s+1}|^2\cos\Delta E_s t + |c_s|^4
    \Big)
\end{equation}
and the corresponding Loschmidt rate is 
\begin{equation}
    \lambda_S(t) = -\frac{1}{N_S}\ln\frac{\mathcal{L}(t)}{\mathcal{L}(0)} \label{E6}
\end{equation}
where it's not appropriate to let $N_S=N=16$ because only five states instead of all eigenstates are 
used to extract the dynamical property. We hence propose using $N_S=\log_2 5$ to calculate $\lambda_S(t)$ where 
$5$ is the number of scarred eigenstates considered.  
We also slightly change the definition of Loschmidt rate to ensure $\lambda_S(0) = 0$ still holds.
Surprisingly, the dynamics of $\lambda_S(t)$ behaves very similar to the 
exact Loschmidt rate $\lambda(t)$ calculated using ED method as shown in Fig.~\ref{dqpt_rate_s}(a).

The Eq.~\eqref{E6} can also be used to explain why the red dots 
(approximate zeros of Loschmidt amplitude) form a sloping line as shown 
in Fig.~\ref{DQPT_rate_complex_plane}. The system size used in this calculation is $N = 12$ 
instead of $16$, as the latter would require significantly more computational time. This is because 
the initial state at $t=0$ has different energy with varying $\tau$, or put it in another way, the overlap 
between $e^{\hat{H}\tau}\ket{\mathbb{Z}_2}$ (not normalized) and eigenstates (especially the scar states) has a 
different peak so the energy level spacing $\Delta E_s$, $\delta$ and the corresponding 
$|c_s|^2$ will change and they certainly will follow the index $s$ of the scar state who 
has the largest overlap with initial state $e^{\hat{H}\tau}\ket{\mathbb{Z}_2}$.
To keep the discussion simple, we assume the ratio between $|c_s|^2$ 
as well as $\delta$ will not change (much) and only consider the effect that 
different $\Delta E_s$ may take. Now we only need to calculate the first zero (besides $t=0$) of 
the derivative of $\mathcal{L}_S(t)$ since it corresponds to the first non-analytical point of $\lambda_S(t)$.
The derivative of $\mathcal{L}_S(t)$ takes the following form
\begin{equation}
    \frac{\mathrm{d}\mathcal{L}_S(t)}{\mathrm{d}t} = -2\sum_{s=p-2}^{p+1}|c_s|^2|c_{s+1}|^2\Delta E_s
    \sin\Delta E_s t
\end{equation}
where we use $p$ to denote the index of scar state who has the largest overlap 
with the initial state $e^{\hat{H}\tau}\ket{\mathbb{Z}_2}$. The relation between 
$\Delta E_p$ and the first, third and fifth zero $\lambda_{S}(t)$ (denoted $t_1, t_2$ and $t_3$ respectively) 
of $\mathrm{d}\mathcal{L}_S(t)/\mathrm{d}t$ is shown in Fig.~\ref{dqpt_rate_s}(b). 

The negative correlation between $t_1,t_2,t_3$ and $\Delta E_p$ and can be clearly seen in 
Fig.~\ref{dqpt_rate_s}(b) and $\Delta E_p$ as well as $p$ also exhibit anticorrelation as shown
in Fig.~\ref{energy_level_spacing}(b). Therefore, in conclusion, as $\tau$ increases, 
the energy of $e^{\hat{H}\tau}\ket{\mathbb{Z}_2}$ will increase and it leads to 
the rise in $p$ which further causes the decrease of $\Delta E_p$ and consequently 
$t_1, t_2, t_3$ increases. Hence for a single branch of red dots, smaller $\tau$ always 
corresponds to smaller $t$ and this can be explained using our approximate Loschmidt rate 
$\lambda_S(t)$ in a phenomenological way.

Due to the similarity between the two-species Rydberg atom array and quantum ANNNI model, we 
believe that the above discussion on the relation between QMBS and DQPT can also 
be applied to the quench dynamics of the quantum ANNNI model, 
i.e., we can use QMBS to explain part of the DQPT in the 
quantum ANNNI model which has already been studied numerically in detail \cite{DQPT2}.

\bibliography{main}

\end{document}